\documentclass[%
 reprint,
superscriptaddress,
amsmath,amssymb,
aps,
pra,
]{revtex4-2}

\usepackage{graphicx}% Include figure files
\graphicspath{{Figures/}}
\usepackage{multirow}
\usepackage{dcolumn}% Align table columns on decimal point
\usepackage{bm}% bold math
\usepackage{bbold}
\usepackage{natbib,hyperref} %references
\usepackage[export]{adjustbox}
\usepackage{wasysym} % circle
\usepackage{simpler-wick} % Wick contractions
\usepackage{mathtools} % Braket notation
\DeclarePairedDelimiter\bra{\langle}{\rvert}
\DeclarePairedDelimiter\ket{\lvert}{\rangle}
\DeclarePairedDelimiterX\braket[2]{\langle}{\rangle}{#1 \delimsize\vert #2}

\usepackage[utf8]{inputenc}
\usepackage[T1]{fontenc}

\usepackage{mathrsfs}
\usepackage{dsfont}

\usepackage{amssymb}
\usepackage{amsmath}
\usepackage{amsfonts}
\usepackage{amsthm}

\usepackage{hyperref}
\hypersetup{pdfpagemode=UseNone}

\newtheorem{theorem}{Theorem}
\newtheorem{definition}[theorem]{Definition}

\newcounter{rem}
\newcommand{\mc}[1]{\mathcal{#1}}
\newcommand{\g}[1]{\mathfrak{#1}}

\def\>{\rangle}
\def\<{\langle}
\newcommand{\proj}[1]{| #1 \rangle\! \langle #1 |}
\newcommand{\idty}{\mathds{1}}
\def\tr{{\rm tr}}
\def\pr{{\rm Pr}}

\def\ii{{\rm i}}
\def\rho{{\varrho}}

\def\textbf#1{{\bf #1}}

\usepackage{xcolor}

 \usepackage{cancel} %Efeito de riscado

\begin{document}

\title{Quantum state inference from coarse-grained descriptions:\\ analysis and an application to quantum thermodynamics}
\author{Ra\'{u}l O.~Vallejos}
\email{vallejos@cbpf.br}
\affiliation{Centro Brasileiro de Pesquisas F\'{\i}sicas (CBPF), Rua Doutor Xavier Sigaud 150, 22290-180 Rio de Janeiro, Brazil}

\author{Pedro Silva Correia}
%\email{pscorreia@id.uff.br}
\affiliation{Instituto de F\'{\i}sica, Universidade Federal Fluminense, Av. Gal. Milton Tavares de Souza s/n, Gragoat\'a 24210-346, Niter\'oi, RJ, Brazil}

\author{Paola Concha Obando}
%\email{pcobando@cbpf.br}
\affiliation{Centro Brasileiro de Pesquisas F\'{\i}sicas (CBPF), Rua Doutor Xavier Sigaud 150, 22290-180 Rio de Janeiro, Brazil}

\author{Nina Machado O'Neill}
%\email{nina@cbpf.br}
\affiliation{Centro Brasileiro de Pesquisas F\'{\i}sicas (CBPF), Rua Doutor Xavier Sigaud 150, 22290-180 Rio de Janeiro, Brazil}

\author{Alexandre Baron Tacla}
%\email{tacla@cbpf.br}
\affiliation{Centro Brasileiro de Pesquisas F\'{\i}sicas (CBPF), Rua Doutor Xavier Sigaud 150, 22290-180 Rio de Janeiro, Brazil}

\author{Fernando de Melo}
\email{fmelo@cbpf.br}
\affiliation{Centro Brasileiro de Pesquisas F\'{\i}sicas (CBPF), Rua Doutor Xavier Sigaud 150, 22290-180 Rio de Janeiro, Brazil}
\date{\today}
\begin{abstract}
	The characterization of physical systems relies on the observable properties which are measured, and how such measurements are performed. Here we analyze two ways of assigning a description to a quantum system assuming that we only have access to coarse-grained properties. More specifically, we compare the Maximum Entropy Principle method, with the Bayesian-inspired recently proposed Average Assignment Map method [P.~S.~Correia \textit{et al}, Phys. Rev. A 103, 052210 (2021)]. Despite the fact that the assigned descriptions by both methods respect the measured constraints, and that they share the same conceptual foundations, the descriptions differ in scenarios that go beyond the traditional system-environment structure. The Average Assignment Map is thus shown to be a more sensible choice for the ever more prevalent scenario of  complex quantum systems. We discuss the physics behind such a difference, and further exploit it in a quantum thermodynamics process.
\end{abstract}

\maketitle

\section{Introduction} 

Whether considering our everyday perception of the surrounding environment or a sophisticated experimental setup, a characterization of a physical 
system is given in terms of measurement results of its observable properties. The characterization of a physical system is thus not unique: besides considering \textit{which} features are being observed, it is also necessary to take into account \textit{how} these features are being observed. 

For the first consideration, the ``which features'' part, given a set of observed properties -- say a set $\mc{O} = \{o_i\}$ of expected values $o_i$  --, the objective of  state inference is to assign to the system a description that abides by the known data. However, more often than not, the number of constraints is not sufficient to single out an unique state for the system, i.e., there are many states which are compatible with the known data. Let $\Omega(\mc{O})$ be the set of all descriptions which are consistent with the knowledge about the system. Back all the way to Laplace's Principle of Indifference, it is only reasonable to appoint the same probability for each description in $\Omega(\mc{O})$, and to assign the average description, $\overline{\Omega(\mc{O})}$, to the system. The use of the Maximum Entropy Principle (MEP)~\cite{jaynes106,jaynes108}, i.e., to assign to the system the description in $\Omega(\mc{O})$ which has the maximum Shannon  entropy, made this intuitive idea more mathematically concrete -- allowing for the inclusion of constraints and symmetries, and extending its reach to continuous sets. The MEP has found applications in fields as diverse as biology, computer science, and financial markets \cite{banavar2010,karmeshu2003,harte2011,boomsma2014,DeMartino2018,Cesari2018,berger1996,bera2008,caticha2014}. Within Physics, the MEP received a great deal of attention as it recovers the ensembles of statistical physics~\cite{pathria}. The same success is observed in the quantum realm, where one now assigns to the system the quantum state that maximizes the von Neumann entropy, while respecting the known constraints \cite{buvzek1998,buvzek2004}.

Concerning ``how''  features are measured, most inference methods, and the MEP in particular, are somewhat reticent. For quantum systems -- which are the subject of the present contribution --, the only point which is systematically taken care of is the locality of the observables. 
Consider for instance the scenarios described by the theory of open quantum systems~\cite{caldeira, breuer2002}. In such a cases it is  possible to split the total system into two parts: the system of interest $S$, which we assume to have access to; and the environment $E$, whose degrees of freedom we have no control of. Besides a possible restriction on the full system (like the total energy), when inferring the total system description, one usually only knows about local properties of the system. 

\begin{figure}[h!]
	\centering
	\begin{tabular}{cc}
	(a) Open quantum system & (b) Generalized coarse-grained scenario \\
		   \includegraphics[width=0.43\linewidth ]{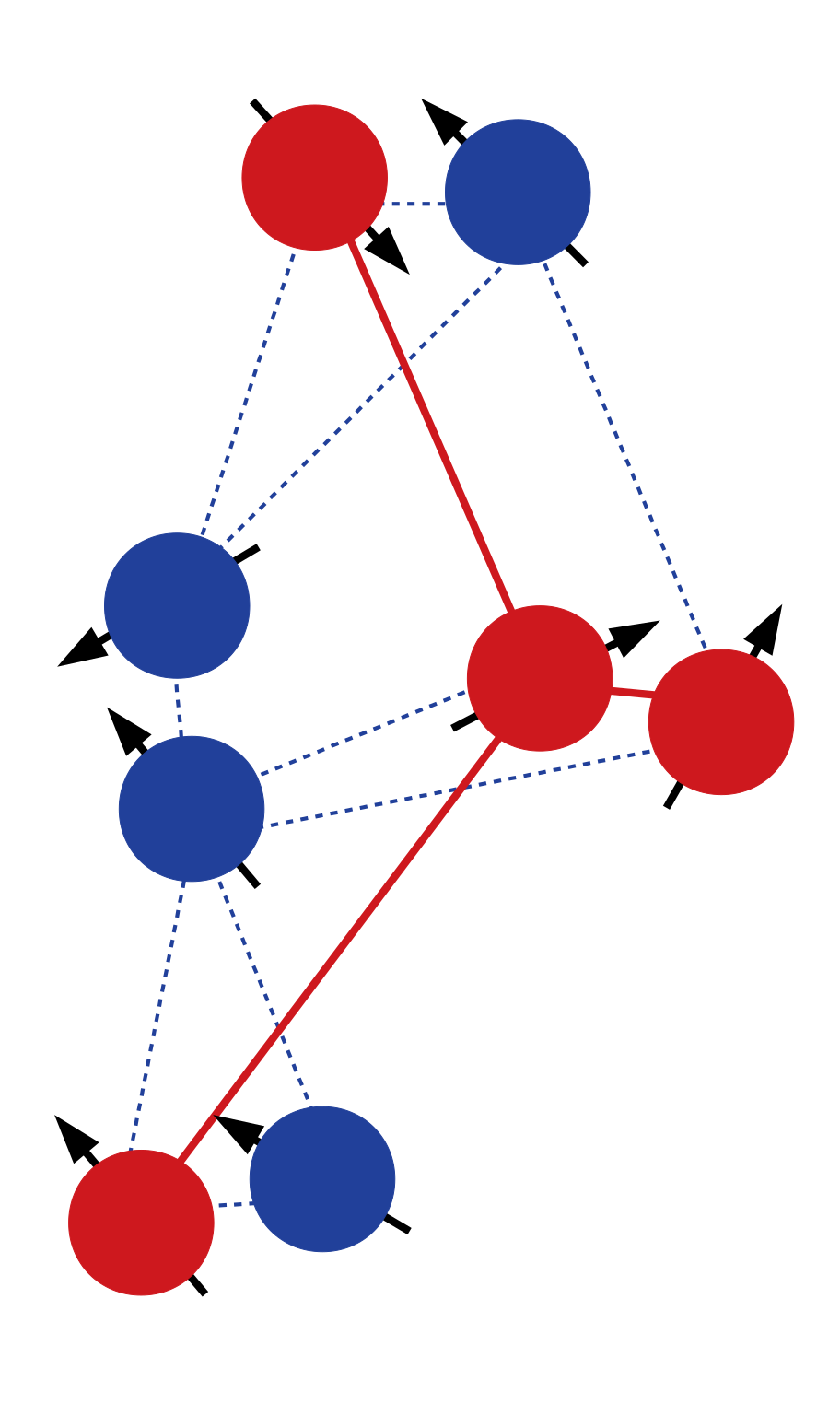} 
		&  \includegraphics[width=0.43\linewidth]{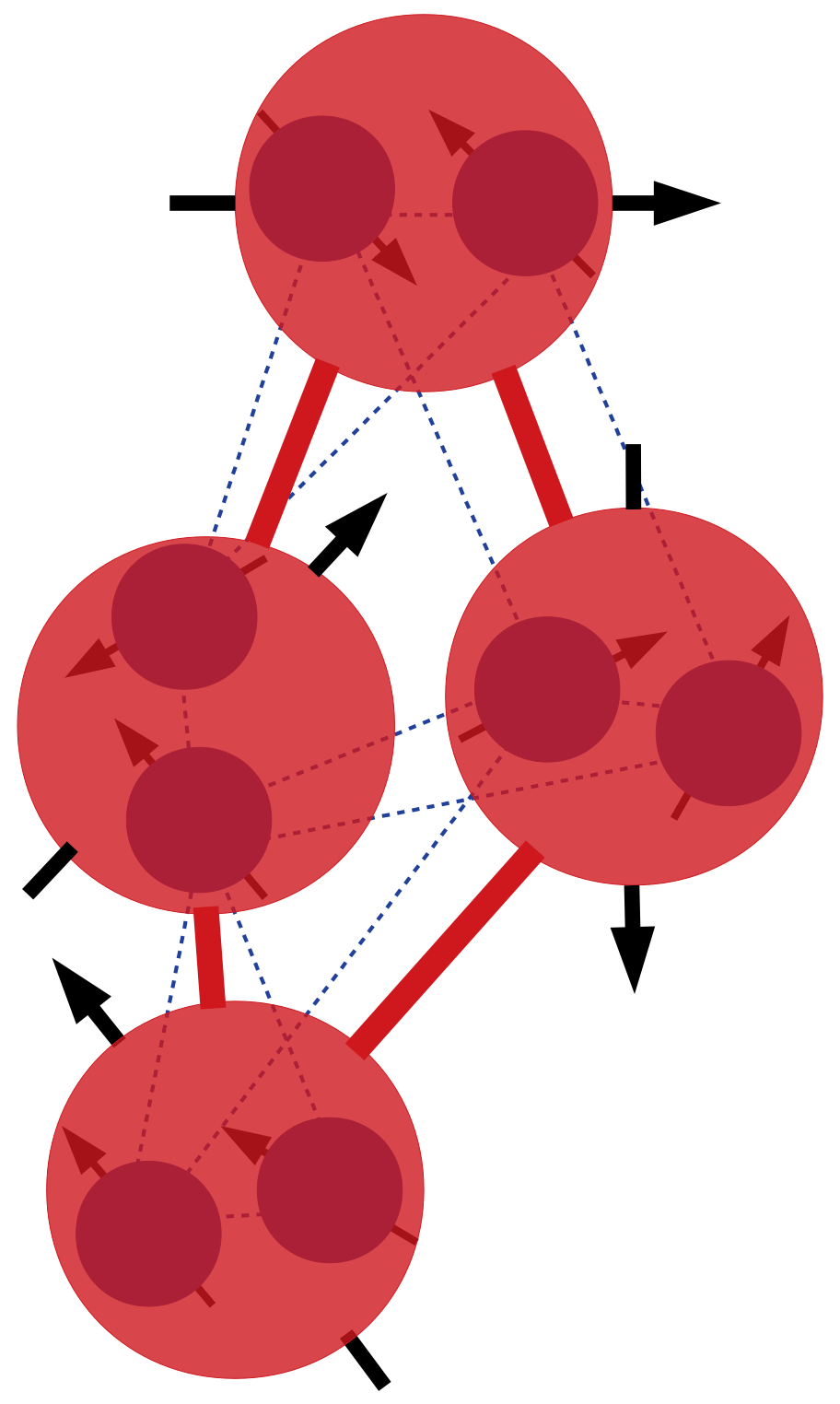}
	\end{tabular}
	\caption{Here we represent a quantum system composed of a set of spins with the interaction among them represented by lines. In a fully microscopic description we must take into account all degrees of freedom (a single spin representing one local degree of freedom). (a) In the open quantum system scenario the effective description of a system is given by splitting the microscopic degrees of freedom between into those we have access to -- the red spins identified as the system --, from those we don’t have access to -- the blue spins identified as the environment – which are removed by the partial-trace map. (b) Example of a scenario where, depending on how coarse-grained is the effective description, the open quantum system scenario cannot be applied. In this case the effective description of the underlying microscopic system is given by effective spins, each one related to a coarse-grained detection of two underlying spins. Physically speaking, such a scenario can model a situation where we do not resolve any individual microscopic degree of freedom of the system, but we only have access to the effective description that emerges from a blurry detection of the whole system.}
	\label{fig:schemeeffective}
\end{figure} 

More concretely,  associate to the total system  a  Hilbert space $\mc{H}_S\otimes\mc{H}_E$, and assume that we know the expected values for a set  $\mc{O}$ of subsystem $S$ properties. Quantum mechanics tells us that there exists a set of observables $\{O_i\}$ and a density matrix $\rho_S$, acting on $\mc{H}_S$, such that $o_i= \tr(O_i \rho_S)$ for all $i\in\{1,\ldots,|\mc{O}|\}$. As the total system lives in $\mc{L}(\mc{H}_S\otimes\mc{H}_E)$ -- with $\mc{L}(\mc{H})$ representing the set of linear operators acting on $\mc{H}$--  the reduced density matrix $\rho_S$ is obtained from the total one, $\rho_{SE}$, by ``tracing out'' system $E$ degrees of freedom, i.e., $\rho_S= \Lambda_{\tr_E}[\rho_{SE}]$. Here, the map $\Lambda_{\tr_E}: \mc{L}(\mc{H}_S\otimes\mc{H}_E) \mapsto \mc{L}(\mc{H}_S)$ is the partial-trace map. If one wants to assign a description to the total system, the local constraints can be extended to the full system as follows:
\begin{align*}
o_i&= \tr(O_i\; \rho_S);\\
&= \tr(O_i \; \Lambda_{tr_E}[\rho_{SE}]);\\
&=\tr(\Lambda^*_{tr_E}[O_i]\; \rho_{SE});\\
&=\tr(O_i\otimes \idty_E \; \rho_{SE}).
\end{align*}
In the above, $\Lambda^*_{tr_E}:\mc{L}(\mc{H}_S)\mapsto\mc{L}(\mc{H}_S\otimes \mc{H}_E)$ is the trace-dual map associated to the partial-trace operation, which only augments the observable with the identity matrix in the $E$ sector. One can now use an inference method to assign a description the total system, which should abide by the  constraints  $o_i=\tr(O_i\otimes \idty_E \; \rho_{SE})$ for all $i\in \{1,\ldots,|\mc{O}|\}$.

However, with the control of quantum systems reaching unprecedented levels, as driven by the development of modern quantum technologies, a new scenario comes forth. In this new picture, highly isolated complex quantum systems are produced whose microscopic description is experimentally challenging -- requiring single particle addressing --, and theoretically hopeless -- due to the exponential increase of system dimension with the number of particles.
In this new scenario, not only the observed data is local, but it also stems from an effective description of the system.

In order to deal with this new class of phenomena, quantum channels with the output dimension smaller than the input dimension, dubbed coarse-graining channels, were used to obtain a system's effective description~\cite{mermin1980, poulin2005, caslavLG, Raeisi2011, Wang2013,Jeong2014, Park2014, cris2017,pedrinho,oleg,cris2019,isadora2020,gabriel2020,cris2020,carlospineda2021}. In this formalism, a coarse-graining map is thus a completely positive map $\Lambda: \mc{L}(\mc{H}_D) \mapsto \mc{L}(\mc{H}_d)$ such that $\dim(\mc{H}_D)>\dim(\mc{H}_d)$. Very much like the partial trace map, the output description has fewer degrees of freedom than the total description. Differently from the partial trace,  general coarse-graining maps do not require a clear-cut split between system and environment, being thus more appropriate to effectively  describe complex highly interacting isolated quantum systems. See Fig.~\ref{fig:schemeeffective} for an example of effective description possible to be characterized by general coarse-graining maps, and how it differs from the usual partial-trace situation.

In the proposed framework, suppose that to a system's effective description we associate a Hilbert space $\mc{H}_d$, and that we know the expected values for a  set $\mc{O}$ of effective properties. As before, there exists a set of observables $O_i$ and a density matrix $\rho_d$, in $\mc{L}(\mc{H}_d)$, such that $o_i= \tr(O_i \rho_d)$ for all $i\in \{1,\ldots, |\mc{O}|\}$. Since $\rho_d$ is an effective description, we can assume that it is the result of a coarse-graining map acting on a microscopic description $\rho_D\in \mc{L}(\mc{H}_D)$, i.e., $\rho_d = \Lambda[\rho_D]$. If we want to infer the microscopic description, we can extend the known constraints as previously:
\begin{align*}
o_i&= \tr(O_i\; \rho_d);\\
&= \tr(O_i \; \Lambda[\rho_{D}]);\\
&=\tr(\Lambda^*[O_i]\; \rho_{D}).
\end{align*}
In the above, $\Lambda^*: \mc{L}(\mc{H}_d) \mapsto \mc{L}(\mc{H}_D)$ is the trace-dual map associated with the coarse-graining channel map $\Lambda$. These constraints can now be included in an inference method to obtain a microscopic description that abides by the coarse-grained data.

While in principle one could use the MEP to assign a description to the microscopic state, when studying the effective dynamics that might emerge from a coarse-grained dynamics~\cite{correia2021}, an assignment map naturally surfaced. Consider the set of all states which are consistent with the coarse-grained data, i.e., the set
\begin{equation}
    \Omega_\Lambda(\mc{O})= \{\psi \in \mc{L}(\mc{H}_D) \; |\; \tr(O_i \Lambda[\psi])=o_i, \; \forall i\in \{1,\ldots,|\mc{O}|\}\}.
    \label{eq:omegaO}
\end{equation}
 As in the foundations of the MEP, in Ref.~\cite{correia2021} it was suggested to assign
to the underlying system the description given by the uniform average of the states, $\overline{\Omega_\Lambda(\mc{O})}^\psi$. The map $\mc{A}_\Lambda:\mc{O}\mapsto \mc{L}(\mc{H}_D)$ that takes the known coarse-grained data and assigns to the system a uniform average among all the fine-grained descriptions that agree with the data was named \textit{Average Assignment Map} (AAM) -- more details below.

Due to the undeniable importance of the Maximum Entropy Principle in the whole Physics, it is the aim of the present contribution to compare the assignments inferred by the  MEP with those inferred  by the AAM. Furthermore, we explore the difference between these two assignment maps in a simple thermodynamical process.

\section{State Inference from Coarse-grained Data}
\label{sec:CGInference}

In this section we give further details about the Average Assignment Map, and recall the Maximum Entropy Principle. For both inference methods, we will assume that we have access to an effective system description, living in $\mc{L}(\mc{H}_d)$, for which we know the coarse-grained data $\mc{O}=\{o_i\}$ coming from the expected values of a set $\{O_i\}$ of corresponding effective observables. Moreover, we will use that the system effective description is generated by a coarse-graining map $\Lambda:\mc{L}(\mc{H}_D)\rightarrow\mc{L}(\mc{H}_d)$.

%In this section we aim to address the opposite direction: we want to define a state inference procedure that maps coarse-grained observed data, described by observables operators with dimension $d$, to a $D$-dimensional fine-grained quantum state, with with $d < D$. We assume  that the coarse-grained description is obtained via a coarse-graining map, say $\Lambda:\mc{L}(\mc{H}_D)\rightarrow\mc{L}(\mc{H}_d)$.  Then, given a physical system described by a set $\mc{O}=\{o_i\}$ of $N_\mc{O}$ effective mean observed data, quantum mechanics assigns coarse-grained observables $O_i \in \mc{L}(\mc{H}_d)$, and microscopic quantum states $\psi:= \proj{\psi}\in \mc{L}(\mc{H}_D)$, such that $o_i= \tr(\Lambda[\psi] O_i)$. 

In the scenario pictured above, notice that a fine-grained state $\psi$ satisfying the coarse-grained constraints $\mathcal{O}$ is in general not unique. The set of all fine-grained states in $\mc{L}(\mc{H}_D)$ that abide by the coarse-grained constraints is then $\Omega_{\Lambda}(\mc{O})$ -- see Eq.\eqref{eq:omegaO}.

\subsection{Average Assignment Map (AAM)}

From an operational perspective, when assembling an effective preparation $\rho$, which is accessed through a coarse-graining map $\Lambda$, and  with properties $\mc{O}$, microscopically we are in fact sampling from the set $\Omega_{\Lambda}(\mc{O})$.  Due to the linearity of the expectation value, this perspective suggests an averaging map $\mc{A}_\Lambda: \mc{O}\rightarrow  \mc{L}(\mc{H}_D)$ that assigns the appropriate description to the microscopic ensemble~\cite{correia2021}: 
\begin{align}
\mathcal{A}_{\Lambda}[\mc{O}]&\equiv\overline{\Omega_{\Lambda}(\mc{O})}^\psi={\int} d{\mu_\psi}\pr_\Lambda(\psi|\mc{O})\,\psi
\label{eq:avgO}
\end{align}
where  $d{\mu_\psi}$ is a prior uniform measure over states in $\mc{L}(\mc{H}_D)$, and  $\pr_\Lambda(\psi|\mc{O})$ is a probability density of having the microscopic state $\psi$ given the macroscopic constraints imposed by $\mc{O}$ and the coarse-graining map $\Lambda$. Note that $\pr_\Lambda(\psi|\mc{O})=0$ for any $\psi \not\in  \Omega_{\Lambda}(\mc{O})$. 

In the particular case where the set $\mc{O}$ is big enough as to allow for the full state reconstruction of $\rho$ in $\mc{L}(\mc{H}_d)$, that is $\mc{O}$ is a tomographically complete set of values, then we can see $\mc{A}_\Lambda$ as a map between states, $\mc{A}_\Lambda: \mc{L}(\mc{H}_d) \rightarrow  \mc{L}(\mc{H}_D)$.
In this case, that without loss of generality we hereafter concentrate on, the AAM map can be more directly written as
\begin{equation}
	\mathcal{A}_{\Lambda}[\rho] = \int d\mu_{\psi} \delta (\Lambda[\psi]-\rho) \, \psi \; .
	\label{eq:ALambda}
\end{equation}
The delta distribution  makes clear that all states in $\Omega_{\Lambda}(\mc{\rho})$ are taken with the same weight in the convex sum.

It is important to discuss the role of the measure $\mu_\psi$. In a situation where the microscopic system is very well isolated, all the ignorance about the system state is classical. In this case, one can consider that the set $\Omega_{\Lambda}(\mc{\rho})$ contains only pure states, and the measure $d\mu_\psi$ will be the Haar measure over pure states in $\mc{H}_D$.  However, if the coupling to the environment cannot be neglected, then one needs to include mixed states in $\Omega_{\Lambda}(\mc{\rho})$, and the choice of the measure $\mu_\psi$ is no longer unique. Below we explore different measures, and see how they change the assigned description. The important point, nonetheless, is to realize that the measure $\mu_\psi$ allows one to include prior knowledge about the way the system is being prepared.

We used two different methods for calculating the integral in \eqref{eq:ALambda} above, the choice of method depending on $\Lambda$ and $\mu_{\psi}$.  The first method consists in direct calculation of the integral using some representation of the $\delta$ (to be described later). 

An alternative method takes advantage of the symmetries of $\Lambda$, and we briefly introduce it in what follows.

\begin{definition}[Channel symmetry] Let $U$ be a unitary operator in $\mc{L}(\mc{H}_D)$.  We say that $U$ is a symmetry of $\Lambda$ if
\begin{equation}
	\Lambda[ U \psi U^\dagger]= \Lambda[\psi]
	\label{eq:def_symm}
\end{equation}
for all $\psi\in\mc{L}(\mc{H}_D)$. 
\end{definition}

It is sufficient to require that identity above holds for all pure states, for
its validity is extended to all linear operators by linearity. Moreover, it is easy to see that all the symmetries of $\Lambda$ form a group.

Assume that $\mu_{\psi}$ is unitarily invariant, i.e.,
\begin{equation}
	\mu_{\psi}= \mu_{ V \psi V^\dagger}
\end{equation}
for all unitary $V$ in $\mc{L}(\mc{H}_D)$. Then, let $U$ be a symmetry of $\Lambda$, and make the change of variables $\psi \to U \psi U^\dagger$ in Eq.\eqref{eq:ALambda}, to obtain
\begin{equation}
	\mathcal{A}_{\Lambda}[\rho] = \int d\mu_{\psi} \delta (\Lambda[\psi]-\rho) \, U \psi U^\dagger \; .
	\label{eq:ALambda1}
\end{equation}
As the assigned state does not change by the application of the symmetry on the fine-grained states, we can average -- with an arbitrary measure --  over a subset of symmetries and still have the same assignment.

Given that averages are linear functions, we can perform the average over the symmetries before the average over the micro-states. In some special cases (we exhibit two examples below) it turns out that such an average over some symmetries presents a remarkable property:
\begin{equation}
 	\wick{\c1 U \psi \c1 U^\dagger} = f(\Lambda[\psi]) \, ,
 	\label{eq:AvUXU}
\end{equation}
with $f$ some function (possibly nonlinear). That is, the average over the symmetries is only a function of the coarse-grained data, as $f(\Lambda[\psi])=f(\rho)$. If that is the case, inserting this result into (\ref{eq:ALambda1}), and using the normalization of the measure, 
we arrive at
\begin{equation}
	\mathcal{A}_{\Lambda}[\rho] =  f(\rho) \; .
	\label{eq:ALambda3}
\end{equation}

\subsection{Maximum Entropy Principle (MEP) assignment}
\label{subsec:MEP}

As mentioned in the introduction, the MEP allows for the inclusion of all sorts of constraints. Let $S(\rho)=- \tr(\rho \ln \rho)$ be the von Neumann entropy of a state $\rho$. The MEP, taken into account the coarse-grained data $\mc{O}$, can be written as:
\begin{align}
        \label{eq:MEP}
    \psi_\text{MEP} =&\arg\max S(\psi)\\
                    & \text{s.t. }\;  \tr(O_i \Lambda[\psi])=o_i \;, \forall  i\in \{1,\ldots, |\mc{O}|\}.\nonumber
\end{align}
The above constraints can be equivalently written as  $ \tr(\Lambda^*[O_i] \psi)=o_i$, where we employed the trace-dual channel associated with the coarse-graining map $\Lambda$.

To solve the MEP optimization, it is the standard practice to use the Lagrangian:

\begin{equation}
\label{eq:lagrangian}
    L= -\tr(\psi\ln{\psi})-\sum_i \lambda_i(\tr(\Lambda^*[O_i] \psi )-o_i),
\end{equation}
where $\lambda_i$ is the Lagrange multiplier associated with the expected value $o_i$. The state that maximizes the entropy, while satisfying all the constraints, can then be written as:
\begin{equation}
\label{eq:psimax}
\psi_{\text{MEP}}= \frac{1}{Z}\exp{(-\sum_i \lambda_i\Lambda^*(O_i))},
\end{equation}
where $Z$ is the partition function, given by $Z= \tr(\exp{(-\sum_i \lambda_i\Lambda^*(O_i))})$, as required by the normalization condition of $\psi_{\text{MEP}}$. Finally, the Lagrange multipliers are related to the constraints via the equations:
\begin{equation}
    o_i = -\frac{\partial}{\partial \lambda_i} \ln Z,
    \label{eq:constraints_MEP}
\end{equation}
for all $i\in\{1,2,\ldots,|\mc{O}|\}$. 

In the case where $\mc{O}$ is sufficient to tomographically reconstruct the effective state $\rho\in \mc{L}(\mc{H}_d)$, one can simply write $\tr(O_i \rho)$  instead of $o_i$ in the expressions above.

\section{AAM vs. MEP: Three Physical Scenarios}
\label{sec:Applications} 

In what follows we compare the AAM and the MEP assignments for three physical scenarios. The first one is the traditional case of an open quantum system with local constraints. For the other two instances the coarse-graining channel cannot be written simply as a partial trace of a subsystem.

\subsection{Partial Trace}

In this scenario the total system is described by a density operator in $\mathcal{L}(\mc{H}_S \otimes \mc{H}_E)$, and the local constraints are described by the system's state $\rho\in \mc{L}(\mc{H}_S)$. The full system's description is thus related to the local one via the partial trace map $\Lambda_{\tr_E}: \mathcal{L}(\mc{H}_S \otimes \mc{H}_E)\mapsto\mc{L}(\mc{H}_S)$. In such a situation, the set of fine-grained states that are consistent with the local constraints is then 
\[ \Omega_{\Lambda_{\tr_E}}(\rho) = \{\psi\in\mc{L}(\mc{H}_{S}\otimes\mc{H}_{E})\; |\; \Lambda_{\tr_E}[\psi]=\rho\}.
\]
\subsubsection{AAM assignment}

As the trace is basis independent,  for any unitary $U_E$ acting on $\mc{H}_E$ we have that $\Lambda_{\tr_E}[(\idty\otimes U_E)\psi (\idty\otimes U_E^\dagger)]=\Lambda_{\tr_E}[\psi]$. The unitary operators of the form $U=\idty\otimes U_E$ thus form a symmetry group for $\Lambda_{\tr_E}$, and we can readily apply the symmetry method to obtain the average assignment for the partial-trace channel.
Performing the average over the symmetry group employing the Haar measure we obtain:
\begin{equation}
	(\wick{ \mathds{1} \otimes \c1 U_E) \, \psi \, (\mathds{1} \otimes \c1 U^\dagger_E })= \Lambda_{\tr_E}[\psi] \otimes \frac{\mathds{1}_E}{d_E} \; ,
	\label{eq:AvUXUtr}
\end{equation}
with $d_E$ the dimension of  $\mc{H}_E$. Following the result in Eq.\eqref{eq:ALambda3}, we conclude that: 
\begin{equation}
	\mathcal{A}_{\tr_E}[\rho] = \rho \otimes \frac{\mathds{1}_E}{d_E} \; .
	\label{eq:AStr}
\end{equation}
Note that $\mu_\psi$ in (\ref{eq:ALambda1}) can be any unitarily invariant measure.
Thus (\ref{eq:AStr}) gives the average assignment for the partial-trace channel, both for averages over 
pure or mixed states.

\subsubsection{MEP assignment}

 Since we are assuming to completely know the system's state, $\rho\in \mc{L}(\mc{H}_S)$, we can consider to know the expectation value of a tomographic set of observables $\{\sigma_i\}$, with $i\in\{x,y,z\}$. For the partial trace map, we then have $\Lambda_{\tr_E}^*(\sigma_i)= \sigma_i\otimes \idty_E$. As such, the state on the total space that maximizes von Neumann's entropy is:
\begin{align}
\psi_{\rm MEP} = &\frac{1}{Z} e^{-\sum_i  \lambda_i\sigma_i\otimes \idty_E},\nonumber\\
=&\frac{ e^{-\sum_i  \lambda_i\sigma_i}}{Z_S} \otimes \frac{\idty_E}{d_E},
\label{eq:chitr1}
\end{align}
with $Z_S=\tr[\exp(-\sum_i  \lambda_i\sigma_i)]$. Nevertheless, as the local system state is fixed by the constraints, we have that the $\lambda_i$ are such that
\begin{equation}
    \psi_{\rm MEP}= \rho \otimes \frac{\idty_E}{d_E}.
    \label{eq:psi_MEP_tr}
\end{equation}

\bigskip

We therefore see that for the partial trace case, where full knowledge about the system' state is given, both the AAM and the MEP assign the same global  description for the total system (system+environment).

\subsection{Blurred and saturated detector}
\label{sec:detector}
Now we introduce our first example of a coarse-graining map that cannot be reduced to  the partial trace of a subsystem. The scenario described by the blurred and saturated coarse-graining map $\Lambda_\text{BnS}$, is suggested by optical-lattices experiments~\cite{gross2017quantum,fig1bloch,fukuhara}. In such  experiments cold atoms are trapped in optical potential wells. Quantum information processing is done by encoding qubits in the atomic energy levels. The read-out of information is commonly carried out by a fluorescence technique~\cite{sherson}, which for our illustrative purposes can be described, in the single atom case, as follows: a laser  with a well-defined frequency is shone over the atom; if the atom is in the excited state, represented by $\ket{1}$, a rapidly decaying transition is resonantly stimulated and  the atom scatters light in all directions; if the atom is the ground state, state $\ket{0}$, the laser is far from resonance, and no light is scattered. 

Now imagine that two neighboring atoms are present in the optical lattice, but due to experimental issues it is not possible to distinguish from which atom the light scattered in the fluorescence measurement comes from. In this situation, the two atoms are seeing as an effective single two-level atom.  The full coarse-graining map modelling this situation is given below (details can be found in~\cite{cris2017,gabriel2020,correia2021,dimolfetta21}):
\begin{equation}
\begin{tabular}{lc|cl}
$\Lambda_\text{BnS}[\ket{00}\bra{00}]
=\ket{0}\bra{0}$ &&& $\Lambda_\text{BnS}[\ket{10}\bra{00}]
=\frac{1}{\sqrt{3}}\ket{1}\bra{0}$ \\

$\Lambda_\text{BnS}[\ket{00}\bra{01}]
=\frac{1}{\sqrt{3}}\ket{0}\bra{1}$ &&& $\Lambda_\text{BnS}[\ket{10}\bra{01}]
=0$ \\

$\Lambda_\text{BnS}[\ket{00}\bra{10}]
=\frac{1}{\sqrt{3}}\ket{0}\bra{1}$ &&& $\Lambda_\text{BnS}[\ket{10}\bra{10}]
=\ket{1}\bra{1}$ \\

$\Lambda_\text{BnS}[\ket{00}\bra{11}]
=\frac{1}{\sqrt{3}}\ket{0}\bra{1}$ &&& $\Lambda_\text{BnS}[\ket{10}\bra{11}]
=0$ \\

$\Lambda_\text{BnS}[\ket{01}\bra{00}]
=\frac{1}{\sqrt{3}}\ket{1}\bra{0}$ &&& $\Lambda_\text{BnS}[\ket{11}\bra{00}]
=\frac{1}{\sqrt{3}}\ket{1}\bra{0}$ \\

$\Lambda_\text{BnS}[\ket{01}\bra{01}]
=\ket{1}\bra{1}$ &&& $\Lambda_\text{BnS}[\ket{11}\bra{01}]
=0$ \\

$\Lambda_\text{BnS}[\ket{01}\bra{10}]
=0$ &&& $\Lambda_\text{BnS}[\ket{11}\bra{10}]
=0$ \\

$\Lambda_\text{BnS}[\ket{01}\bra{11}]
=0$ &&& $\Lambda_\text{BnS}[\ket{11}\bra{11}]
=\ket{1}\bra{1}$
\label{eq:CGblurredmap}
\end{tabular}
\end{equation}

Although inspired by the resonance fluorescence measurement of cold atoms in optical lattices, the above map has no intention to fully capture all the experimental nuances. Nevertheless, it is worth stressing that $\Lambda_\text{BnS}$ cannot be reduced to a simple partial trace of one of the subsystems. This can be immediately apprehended by observing the diagonal elements $\Lambda_\text{BnS}[\proj{01}]=\Lambda_\text{BnS}[\proj{10}]=\proj{1}$ -- the equalities here are not compatible with the partial trace of either subsystem. 

We then proceed to contrast the AAM and the MEP assignments for this, physically inspired, but not traditionally described situation. As for the partial trace case, we will assume to have a tomographic description of the effective system, i.e., we will assume that to know $\rho\in\mc{L}(\mc{H}_2)$, and we want to determine an assignment in $\mc{L}(\mc{H}_4)$ which abide by the constraints. The set of fine-grained states that are compatible with the effective system is now:
\[ \Omega_{\Lambda_{\text{BnS}}}(\rho) = \{\psi\in\mc{L}(\mc{H}_4)\; |\; \Lambda_\text{BnS}[\psi]=\rho\}.
\]

\subsubsection{AAM assignment}

In what follows we obtain the AAM assignment considering both pure and mixed states in the set $\Omega_\text{BnS}(\rho)$. We start by employing the symmetry method for pure states, and then continue with direct integration for both cases.

\paragraph{Symmetry method -- pure-state measure.}

Let $\psi=[\psi]_{ij}$, with $i,j\in\{0,1,2,3\}$, be the matrix representation in the computational basis of a generic state in $\mc{L}(\mc{H}_4)$ -- where for the moment we ignore the trace, positivity, and purity constraints that a pure quantum state must fulfill. For the present coarse-graining map, we have
\begin{equation}
	\Lambda_\text{BnS}[\psi] =
	\begin{pmatrix}
    \psi_{00}                       & \dfrac{\psi_{01} + \psi_{02} + \psi_{03}}{\sqrt{3}} \\
\dfrac{\psi_{10} + \psi_{20} + \psi_{30}}{\sqrt{3}}    &  \psi_{11} + \psi_{22} + \psi_{33}
    \end{pmatrix}		\; .
	\label{eq:Lam_chi_Det} 
\end{equation}
Using the above expression in the definition of a channel symmetry Eq.\eqref{eq:def_symm},  as we show in Appendix \ref{app1},  the symmetries of $\Lambda_\text{BnS}$ can be written as 
\[
U= \idty\oplus \idty\oplus V,
\]
where the operators on the RHS act on the following subspaces:
the first subspace is given by the $\text{span}\{\ket{0}\}$, the second one by $\text{span}\{(0,1,1,1)^T/\sqrt{3}\}$, and the third one is an arbitrary two-dimensional subspace orthogonal to the first two. In this way, $V$ is an arbitrary unitary matrix in two dimensions, such that the group of symmetries of $\Lambda_\text{BnS}$ is isomorphic to $U(2)$.
Following the recipe of the symmetry method explained above, we chose a parameterization of $V\in U(2)$, and average over the symmetry group to obtain:
\begin{equation}
	\wick{\c1 U \psi \c1 U^\dagger} =  
	\begin{pmatrix}
		\Circle        & \triangle & \triangle & \triangle \\
		\triangle^\ast & \Diamond  & \square   & \square\\
		\triangle^\ast & \square   & \Diamond  & \square \\
		\triangle^\ast & \square   & \square   & \Diamond \\
	\end{pmatrix} \; ,
	\label{eq:AvUXU_Det}
\end{equation}
where the symbols are  given by
\begin{align}
	\Circle    & = \psi_{00} \; , \\
	\triangle  & = \frac{1}{3} (\psi_{01}+\psi_{02}+\psi_{03}) \; , \\
	\Diamond   & = \frac{1}{3} (\psi_{11}+\psi_{22}+\psi_{33}) \; , \\
	\square    & = \frac{1}{6} (\psi_{12}+\psi_{13}+\psi_{23}+
	                            \psi_{21}+\psi_{31}+\psi_{32}) \; .
\end{align}
Looking at Eq.\eqref{eq:Lam_chi_Det}, and imposing $\Lambda_\text{BnS} [\psi]=\rho$, we note that for an arbitrary $\psi$ we must have:
\begin{align}
	\Circle    & = \rho_{00} \; , &
	\Diamond   & = \frac{\rho_{11}}{3} \; , &
	\triangle  & = \frac{\rho_{01}}{\sqrt{3}} \;,
	\label{eq:Circ_Dia_tri}
\end{align}
where $\rho_{mn}$, with $m,n\in\{0,1\}$, are the matrix coefficients of $\rho$ in the computational basis.

The $\square$, however, cannot be directly written only in terms of the components of $\rho$. If we now assume that $\psi$ is a pure state, then we can write:
\begin{equation}
\psi_{ij} = c_i c_j^\ast \; , \text{where } \ket{\psi}=(c_0,c_1,c_2,c_3)^T \; .
\label{eq:chi_pur}
\end{equation}
In this case we have
\begin{equation}
\square = \frac{3\; |\triangle|^2}{2 \;\Circle}-\frac{\Diamond}{2} = 
          \frac{|\rho_{01}|^2}{2 \;\rho_{00}}-\frac{\rho_{11}}{6}     \; .
\label{eq:square}
\end{equation}
Putting all together, when $\mu_{\psi}$ is the uniform measure over pure states, then the  average assignment map leads to:
\footnotesize
\begin{equation}
\mathcal{A}_{\Lambda_\text{BnS}}[\rho]=\\\begin{pmatrix}
\rho_{00} & \dfrac{\rho_{01}}{\sqrt{3}} & \dfrac{\rho_{01}}{\sqrt{3}} & \dfrac{\rho_{01}}{\sqrt{3}} \\
\dfrac{\rho_{01}^\ast}{\sqrt{3}} & \dfrac{\rho_{11}}{3} & \dfrac{|\rho_{01}|^2}{2\rho_{00}}-\dfrac{\rho_{11}}{6} & \dfrac{|\rho_{01}|^2}{2\rho_{00}}-\dfrac{\rho_{11}}{6}\\
\dfrac{\rho_{01}^\ast}{\sqrt{3}} & \dfrac{|\rho_{01}|^2}{2\rho_{00}}-\dfrac{\rho_{11}}{6} & \dfrac{\rho_{11}}{3} & \dfrac{|\rho_{01}|^2}{2\rho_{00}}-\dfrac{\rho_{11}}{6} \\
\dfrac{\rho_{01}^\ast}{\sqrt{3}} & \dfrac{|\rho_{01}|^2}{2\rho_{00}}-\dfrac{\rho_{11}}{6} & \dfrac{|\rho_{01}|^2}{2\rho_{00}}-\dfrac{\rho_{11}}{6} & \dfrac{\rho_{11}}{3}
\end{pmatrix}.
\label{eq:avblurred}
\end{equation}
\normalsize
It is interesting to realize that in this case the AAM depends in  a non-linear way on the elements of $\rho$. This non-linearity was discussed in greater detail, and applied in an effective communication scenario, in Ref.~\cite{pedrinho}.

\paragraph{Explicit integration method -- pure-state measure.}

Before using direct integration to calculate the matrix elements corresponding to  $\square$ 
in Eq.\eqref{eq:AvUXU_Det} for the case of a mixed-state measure, first we reproduce the pure-state measure result. This will allow us to exhibit the method in simpler setting. After that, the generalization to a mixed-state measure will be relatively straightforward.

In what follows, in order to simplify the calculations' presentation, we employ a hybrid notation for the effective states $\rho\in\mc{L}(\mc{H}_2)$: in part we use its computational basis elements, and in part its Bloch vector representation. Given a fixed effective state $\rho$ we write it as:
\begin{equation}
\label{eq:hybrid}
\rho = \begin{pmatrix}
        \rho_{00}&0\\
        0&\rho_{11}
\end{pmatrix} + \frac{1}{2}\big(x\; \sigma_x +y\; \sigma_y\big).
\end{equation}
In the above expression, $\sigma_i$, with $i\in\{x,y,z\}$, are the usual Pauli matrices, $x=\tr(\rho\, \sigma_x)=\frac{1}{2} \text{Re}(\rho_{01}) $, and $y=\tr(\rho\, \sigma_y)= \frac{1}{2} \text{Im}(\rho_{01})$.

As we are performing the integration over the pure-state measure (the uniform Haar measure), we directly make use of the pure state structure, as specified in Eq.\eqref{eq:chi_pur}. The action of the coarse-graining map $\Lambda_\text{BnS}$ in a generic pure state can then be explicitly written as:
\begin{equation}
	\Lambda_\text{BnS}[\psi] =
	\begin{pmatrix}
    c_0^\ast c_0                      &  c_0 (c_1^\ast + c_2^\ast + c_3^\ast)/\sqrt{3} \\
  c_0^\ast (c_1 + c_2 + c_3)/\sqrt{3} &  c_1^\ast c_1 + c_2^\ast c_2+ c_3^\ast c_3
    \end{pmatrix}		\; .
	\label{eq:Lam_chi_pur} 
\end{equation}
With the above representations, the average assignment map, Eq.\eqref{eq:ALambda}, can be expanded as follows for the present case:
\begin{align}
	\mathcal{A}_{\Lambda_\text{BnS}}[\rho] \propto \int d\mu_{\psi} 
	        & \delta \big(c_0^\ast c_0-\rho_{00}\big)  \times \nonumber\\
			& \delta \big(c_1^\ast c_1 + c_2^\ast c_2+ c_3^\ast c_3-\rho_{11}\big)  \times \nonumber\\
			& \delta \big(\tr(\Lambda_\text{BnS}[\psi]\, \sigma_x)- x\big)\times \nonumber    \\  
			& \delta \big(\tr(\Lambda_\text{BnS}[\psi]\, \sigma_y)- y\big) \,\psi   \equiv \mathcal{I }\; ,
\label{eq:int_del_0}    
\end{align}
where $\propto$ indicates that the average state given by the l.h.s, expression, named $\mc{I}$,  may not be normalized. The normalization constant $\mathcal{N}$ is given by 

\begin{align}
	\mathcal{N}\equiv \int d\mu_{\psi} 
	        & \delta \big(c_0^\ast c_0-\rho_{00}\big)  \times \nonumber\\
			& \delta \big(c_1^\ast c_1 + c_2^\ast c_2+ c_3^\ast c_3-\rho_{11}\big)  \times \nonumber\\
			& \delta \big(\tr(\Lambda_\text{BnS}[\psi]\, \sigma_x)- x\big)\times \nonumber    \\  
			& \delta \big(\tr(\Lambda_\text{BnS}[\psi]\, \sigma_y)- y\big) \; .
\label{eq:int_del_N}			
\end{align}
The average assignment will then be given as a quotient of two integrals
\begin{equation}
	\mathcal{A}_\text{BnS}[\rho] = \frac{\mathcal{I}}{\mathcal{N}} \; .
\end{equation}
Both integrals will be evaluated using the same scheme we now show. The first step is to write the integration measure as a function of the coefficients of $\ket{\psi}=(c_0,c_1,c_2,c_3)^T\equiv c$:
\begin{equation}
	d\mu_{\psi} = \prod_i d \text{Re}(c_i) d \text{Im}(c_i) \equiv d(c^\dagger,c).
	\label{eq:mes_cc} 
\end{equation}
 To write the measure as $d(c^\dagger,c)$ is a reminder of its 
 explicit dependence on the complex coefficients $c_i$.
% %
Note that the normalization of $\ket{\psi}$ is implicit in the first two deltas in (\ref{eq:int_del_0}), as long as  $\rho_{00}+\rho_{11}=1$.

For the present scenario, the effective state $\rho$ is fixed, and therefore its components in the computational basis, $\rho_{ij}$, are also fixed. Nevertheless, for the sake of calculation we will take them as independent variables. As we are going to perform integral transformations, more specifically Laplace ($\g{L}$) and Fourier ($\g{F}$) transformations \cite{laliena99,laksh08}, and then take their inverse transformations, their ``temporary'' independence will be just a calculation artifact. Within this perspective, as $\rho_{00}$ and $\rho_{11}$ are positive valued, we can write:
\begin{align}
    \delta(c_0^*c_0 -& \rho_{00})=\nonumber\\
    &=\g{L}^{-1}\{\g{L}\{\delta(c_0^*c_0 - \rho_{00})\}(s_0)\}(\rho_{00});\nonumber\\
    &=\g{L}^{-1}\left\{\int_0^\infty d\rho_{00}\,e^{-s_0 \rho_{00}} \delta (c_0^*c_0-\rho_{00})\right\}(\rho_{00});\nonumber\\
    &=\g{L}^{-1}\left\{e^{-s_0\,c_0^*c_0}\right\}(\rho_{00}).
\end{align}
Similarly, for $\rho_{11}$:
\begin{align}
    \delta(c_1^\ast c_1 + c_2^\ast c_2&+ c_3^\ast c_3-\rho_{11})=\nonumber\\
    &=\g{L}^{-1}\left\{e^{-s_1\,(c_1^\ast c_1 + c_2^\ast c_2+ c_3^\ast c_3)}\right\}(\rho_{11}).
\end{align}
 
 For the remaining components, $x$ and $y$, as they might assume negative values, we perform Fourier transformations:
 \begin{align}
     \delta \big(\tr(&\Lambda_\text{BnS}[\psi]\, \sigma_x)- x\big)=\nonumber\\
     &=\g{F}^{-1}\{\g{F}\{\delta \big(\tr(\Lambda_\text{BnS}[\psi]\, \sigma_x)- x\big)\}(k_x)\}(x);\nonumber\\
     &=\g{F}^{-1}\left\{\int_{-\infty}^\infty dx\,e^{- \ii k_x\,x}\delta\big(\tr(\Lambda_\text{BnS}[\psi]\, \sigma_x)- x\big)\right\}(x);\nonumber\\
     &=\g{F}^{-1}\left\{e^{- \ii k_x\,\tr(\Lambda_\text{BnS}[\psi]\, \sigma_x)}\right\}(x);\nonumber\\
     &=\g{F}^{-1}\left\{e^{- \ii k_x \, c\, \Lambda_\text{BnS}^*[\sigma_x]\,c^\dagger)}\right\}(x).
 \end{align}
 In the last line we employed the dual channel $\Lambda_\text{BnS}^*$, the definition of $\psi=c c^\dagger$, and the ciclic property of the trace.
 Proceeding exactly in the same fashion for the $y$ component, we get:
 \begin{equation}
     \delta \big(\tr(\Lambda_\text{BnS}[\psi]\, \sigma_y)- y\big)=\g{F}^{-1}\left\{e^{- \ii k_y \, c\, \Lambda_\text{BnS}^*[\sigma_y]\,c^\dagger)}\right\}(y).
 \end{equation}

After transforming all the deltas as shown above, and interchanging the order of integration between the inverse transformations and the 
integral over states, we can define the the  Laplace/Fourier transformed versions of  both $\mathcal{I}$ and $\mathcal{N}$:%
\begin{align}
	\tilde{\mathcal{I}}[s_0,s_1,k_x,k_y]  & =
	 \int d(c^\dagger,c) \, e^{-c^\dagger A c} \, c \, c^\dagger  \; , \\
	\tilde{\mathcal{N}}[s_0,s_1,k_x,k_y]  & =
	 \int d(c^\dagger,c) \, e^{-c^\dagger A c}                    \; , 
	\label{eq:gauss_int}
\end{align}
where the matrix $A$ is given by
\begin{align}
	A & = {\rm diag} \, (s_0,s_1,s_1,s_1) + \ii \Lambda^\ast[k_x \sigma_x + k_y \sigma_y]  \\
	  & = \begin{pmatrix}
	           s_0 & \frac{\ii k_x + k_y}{\sqrt{3}} & \frac{\ii k_x + k_y}{\sqrt{3}} & \frac{\ii k_x + k_y}{\sqrt{3}} \\
                \frac{\ii k_x - k_y}{\sqrt{3}} & s_1 & 0   & 0   \\
               \frac{\ii k_x - k_y}{\sqrt{3}} & 0   & s_1 & 0   \\
               \frac{\ii k_x - k_y}{\sqrt{3}} & 0   & 0   & s_1 \\
\end{pmatrix}.
\end{align}
In this way, the integrals for $\tilde{\mc{I}}$ and $\tilde{\mc{N}}$ are well defined Gaussian integrals (as $s_0$ and $s_1$ are positive, $A+A^\dagger$ is positive definite), which can be readily performed~\cite{altland}:
\begin{align}
	\tilde{\mathcal{N}}[s_0,s_1,k_x,k_y]  & = \dfrac{\pi^4}{\det A}  \; , 
	\label{eq:N_det_pure} \\
	\tilde{\mathcal{I}}_{ij} = \int d(c^\dagger,c) \, e^{-c^\dagger A c} c_i^\ast c_j & = 
	   \dfrac{\pi^4}{\det A} A^{-1}_{ji} \; ,
	\label{eq:I_det_pure}
\end{align}
where $A^{-1}_{ji}$ stands for the $ji$ element of $A^{-1}$.

The next step is to perform the inverse transforms. 
Although tedious, this step can be directly performed 
with the aid of a mathematical software like Mathematica~\cite{Mathematica}. Further details are to be found in the Appendix~\ref{app2}.

With this final step, all the elements of $\mc{A}_{\Lambda_\text{BnS}}(\rho)$, and in particular the one denoted by the $\square$, can be obtained. The result, as expected, is exactly the same as the one shown in Eq.\eqref{eq:avblurred}.

\paragraph{Explicit integration method -- mixed-state measure.}

Up to now, when dealing with the $\Lambda_\text{BnS}$ coarse-graining map, we assumed that the set $\Omega_{\Lambda_\text{BnS}}(\rho)$ contained only pure states. Now we deal with the situation where the  fine-grained system might be coupled to some other system that we do not have control of, some environment $E$, and thus all the preparations of our system of interest in $\mc{L}(\mc{H}_D)$ are already  intrinsically  described by a mixed state.  In such a situation, there exists no unique uniform measure over mixed states. One way to circumvent this is by employing the unique measure of pure states over the whole space, system and environment, and trace-out the environmental degrees of freedom\cite{zyczkowski01}. 

Within this mindset, we write:
\[
\Omega_{\Lambda_\text{BnS}}(\rho)=\{\tr_E(\Psi)\;|\; \Psi \in \mc{L}(\mc{H}_D\otimes\mc{H}_E), \; \Lambda_\text{BnS}\circ\tr_E[\Psi]=\rho\},
\]
with $\Psi$ a pure state in $\mc{L}(\mc{H}_D\otimes\mc{H}_E)$. The Average Assignment Map is then expressed as:
\begin{equation}
	\mathcal{A}_{\Lambda_\text{BnS}}[\rho] = \int d\mu_{\Psi} \delta (\Lambda\circ\tr_E[\Psi]-\rho) \, \tr_E[\Psi],
	\label{eq:ALambda_mixed}
\end{equation}
with $d \mu_\Psi$ the Haar measure over $\mc{H}_D\otimes\mc{H}_E$. It is important to notice that different environment dimensions will lead to different induced measures of mixed states in $\mc{L}(\mc{H}_D)$~\cite{zyczkowski01}. This is physically related to ``how'' open our system of interest is, and its implication will be explored below.

Let $\{\ket{m}\}$, with $m\in\{0,1,2,3\}$, be the computational basis in $\mc{H}_D$, and let $\{\ket{\phi_k}\}$, with $k\in\{1,2,\ldots,d_E\}$, be an orthonormal basis for $\mc{H}_E$,  whose dimension is $d_E$. Then, in order to follow the steps for the pure-measure case, we write:
\begin{equation}
	|\Psi \rangle = \sum_{k=1}^{d_E} \sum_{m=0}^3 c_m^k\ket{m}_S \otimes |\phi_k \rangle_E  \; 
\end{equation}
and evaluate
\begin{align}
   \tr_E \proj{\Psi}=& \nonumber\\
	    & = \sum_{k=1}^{d_E} 
	\left( \sum_{m=0}^3 c_m^k \ket{m} \right) \left( \sum_{m'=0}^3 c_{m'}^{k*}\bra{m'} \right)\nonumber \\
		& \equiv \sum_{k=1}^{d_E} \proj{\psi^{k}}.
\end{align}
In the above, 
\begin{equation}
	|\psi^k\rangle =  \sum_{m=0}^3 c_{m}^k |m \rangle
\end{equation}
are sub-normalized pure states in $\mc{H}_D$.
Associating  to each pure state $|\psi^k\rangle$ a coordinate vector, i.e.,  $|\psi^k \rangle \to c^k=(c_0^k,c_1^k,c_2^k,c_3^k)^T$ we can write $\tr_E\proj{\Psi}= c^kc^{k\dagger}$, where the summation over repeated indices is assumed whenever it is not explicitly written.

Like in the pure-state measure, we can split the deltas and define:
\begin{align}
	\mathcal{A}_{\Lambda_\text{BnS}}[\rho] \propto \int d\mu_{\Psi} 
	        & \delta \big(c_0^{k \ast} c_0^k-\rho_{00}\big)  \times \nonumber\\
			& \delta \big(c_1^{k \ast} c_1^k + c_2^{k \ast} c_2^k+ c_3^{k \ast} c_3^k-\rho_{11}\big)  \times \nonumber\\
			& \delta \big(c^{k\dagger}\, \Lambda_\text{BnS}^*[\sigma_x]c^k- x\big)\times \nonumber    \\  
			& \delta \big(c^{k\dagger}\, \Lambda_\text{BnS}^*[\sigma_y]c^k- y\big) \,c^{k^\prime}c^{k^\prime\dagger}   \equiv \mathcal{I_\text{mixed} }\; ,
\label{eq:int_del_mix}    
\end{align}
\begin{align}
	\mathcal{N}_{\text{mixed}}\equiv  \int d\mu_{\Psi} 
	        & \delta \big(c_0^{k \ast} c_0^k-\rho_{00}\big)  \times \nonumber\\
			& \delta \big(c_1^{k \ast} c_1^k + c_2^{k \ast} c_2^k+ c_3^{k \ast} c_3^k-\rho_{11}\big)  \times \nonumber\\
			& \delta \big(c^{k\dagger}\, \Lambda_\text{BnS}^*[\sigma_x]c^k- x\big)\times \nonumber    \\  
			& \delta \big(c^{k\dagger}\, \Lambda_\text{BnS}^*[\sigma_y]c^k- y\big),
\label{eq:norm_mix}    
\end{align}
where the integration measure now reads 
\begin{equation}
	d\mu_{\Psi} = \prod_{k=1}^{d_E}  d(c^{k \dagger},c^k).
\end{equation}
Following the pure-state measure recipe, we now perform the integral transformations to obtain:
\begin{align}
	\tilde{\mathcal{I}}_{\rm mixed}  & =\sum_{k^\prime=1}^{d_E}\left(\prod_{k=1}^{d_E} \int d(c^{k\dagger},c^k) \, e^{-c^{k\dagger} A c^k}\right) \, c^{k^\prime} \, c^{k^\prime\dagger} \nonumber \\
        & = d_E \, \tilde{\mathcal{N}}^{d_E-1} \, \tilde{\mathcal{I}}[s_0,s_1,k_x,k_y] \; , 
    \label{eq:I_det_mix}
\end{align}
and 
\begin{align}
	\tilde{\mathcal{N}}_{\rm mixed}  & =\left(\prod_{k=1}^{d_E} \int d(c^{k\dagger},c^k) \, e^{-c^{k\dagger} A c^k}\right) \\
	                                 & = \tilde{\mathcal{N}}^{d_E}[s_0,s_1,k_x,k_y] \; , 
	\label{eq:N_det_mix}                            
\end{align}
with $A$, $\tilde{\mathcal{N}}$ and $\tilde{\mathcal{I}}$ as in the previous section (pure case).

After doing the inverse transforms (see Appendix~\ref{app2}), we arrive at the following result:
\begin{equation}
	\square= 
	\begin{cases*} 
	   \dfrac{d_E}{3 d_E-1}  \dfrac{|\rho_{01}|^2}{  \rho_{00}} - \dfrac{\rho_{11}}{3(3 d_E-1)}  & \text{mixed states} \\[7pt]
	   \dfrac{                      |\rho_{01}|^2}{2 \rho_{00}} - \dfrac{\rho_{11}}{6}           & \text{pure states}  
	\end{cases*} \; .
\label{eq:ass_det_mix_pure}
\end{equation}
We included the expression for pure states for the sake of comparison.
The pure case corresponds to $d_E=1$.

\begin{figure}
\includegraphics[width=\linewidth]{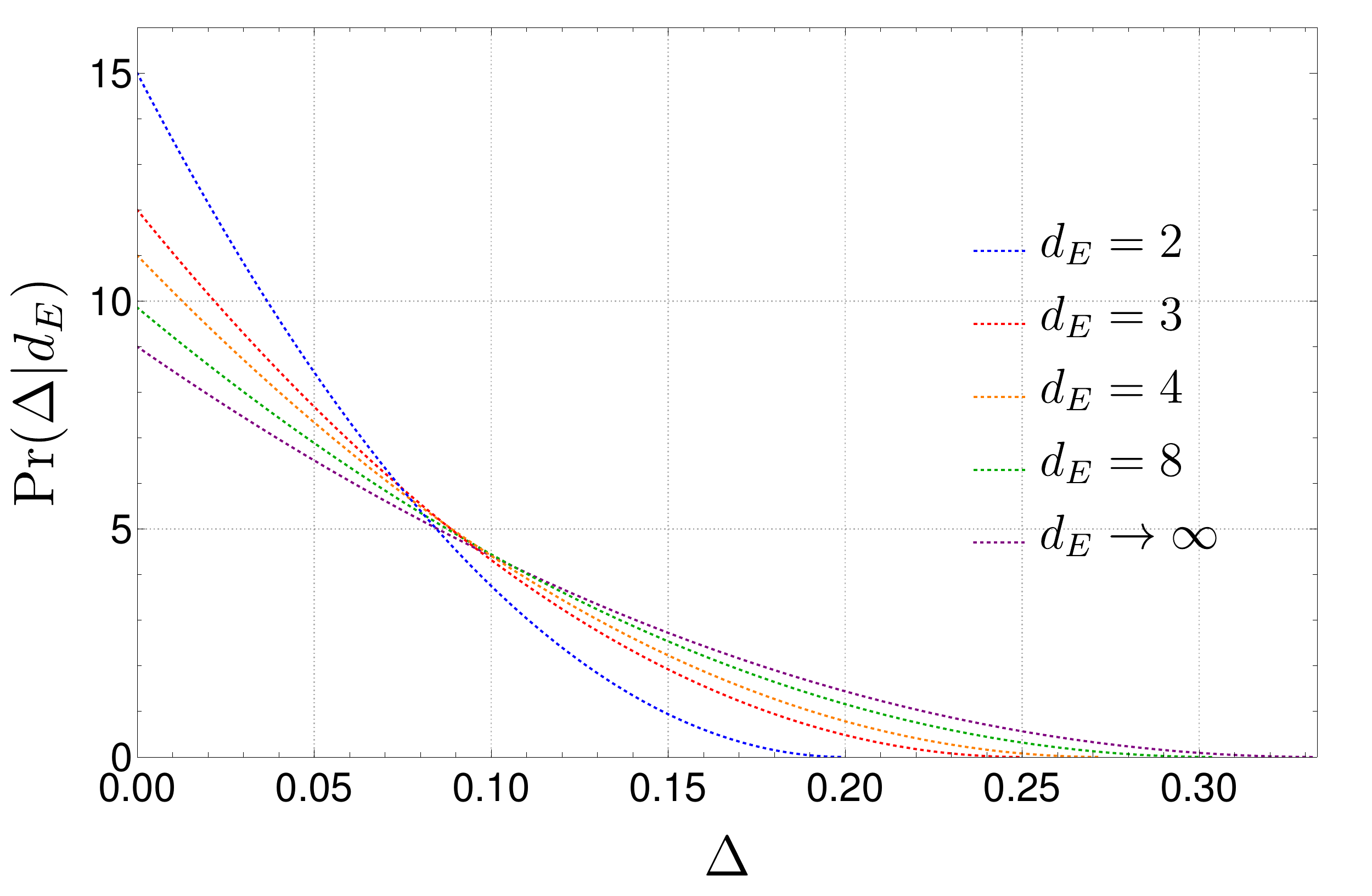}
\caption{\textbf{Comparison between pure and mixed priors.} 
	The curves represent  the probability density $\pr(\Delta|d_E)$ of obtaining a distance $\Delta$ between the pure and mixed assignments, given a fixed environment dimension $d_E$, when sampling effective states uniformly from the Bloch sphere. According to their probabilities at the origin, the curves correspond to $d_E=2,3,4,8,$ and $\infty$ (from top to bottom).}
\label{fig:pofd}
\end{figure}

In Fig.~\ref{fig:pofd} we show a comparison between pure- and mixed-prior assignments 
using the trace distance ($\Delta$) as a measure of similarity.
We plot the probability density $\pr(\Delta|d_E)$ of obtaining a distance $\Delta$ between the assignments using the pure-state measure and the mixed-state measure, given a fixed environment dimension $d_E$. An analytical expression for $\pr(\Delta|d_E)$ is derived in Appendix~\ref{app2_trace}, and reads:
\begin{equation}
\pr(\Delta| d_E) = \frac{3 (\Delta -2 a)^2}{8 a^3}, \qquad 0 \le \Delta \le 2a \; ,
\label{eq:p_of_delta}
\end{equation}
where
\begin{equation}
a = \frac{d_E-1} {2 (3 d_E-1)    }. 
\end{equation}
As expected, and clearly shown in Fig.~\ref{fig:pofd}, the probability of large differences between the pure and mixed measures is greater for big environments. In other words, the error that one makes in assuming a closed system, and thus employing a pure-sate measure as prior, is smaller when the environment is  small.

\subsubsection{MEP assignment}

To obtain the traditional MEP assignment for the situation described by the Blurred and Saturated Detector coarse-graining map, $\Lambda_\text{BnS}$, we start from the general MEP assignment prescription:
\begin{equation}
\psi_{\rm MEP}^\rho= \frac{1}{Z} \exp{\left(-\sum_{i\in\{x,y,z\}} \lambda_i\Lambda_\text{BnS}^*[\sigma_i]\right)},
\label{eq:mep_bns1}
\end{equation}
with $\Lambda_\text{BnS}^*:\mc{L}(\mc{H}_2)\rightarrow \mc{L}(\mc{H}_4)$ the trace-dual map related to $\Lambda_\text{BnS}$. The matrix in the exponent of Eq.\eqref{eq:mep_bns1} can be exactly diagonalized, and then exponentiated. 

The $\psi_\text{MEP}$ for the present scenario has the same structure shown in Eq.\eqref{eq:AvUXU_Det}, but now its elements read:

\begin{align}
\Circle    & =  \dfrac{1}{Z}
        \left(\cosh \lambda - \dfrac{\lambda_z \sinh{\lambda}}{\lambda} \right) \; , \\[5pt]
\triangle  & = -\dfrac{1}{Z}
        \left(\frac{(\lambda_x-i \lambda_y) \sinh \lambda }{\sqrt{3} \, \lambda } 
                                                                        \right) \; , \\[5pt]
\square    & = \dfrac{1}{3Z}
        \left( \frac{\lambda_z \sinh \lambda }{\lambda } - e^{\lambda_z} +\cosh \lambda  
                                                                        \right) \; , \\[5pt]
\Diamond   & =  \dfrac{1-\Circle}{3}  \;,
\end{align}
with the partition function given by
\begin{equation}
 Z= 2 \left( \cosh{\lambda} + e^{\lambda_z}\right).
\end{equation}
By imposing the constraint
\begin{equation}
\Lambda[\psi_{\rm MEP}^\rho]= \rho \; ,
\label{eq:constraints_MEP_BnS}
\end{equation}
we find that the elements denoted by $\Circle$, $\Diamond$, and $\triangle$ have 
the same expression as in the case of the average assignment [see Eq.~(\ref{eq:Circ_Dia_tri})].
However, the elements expressed by a $\square$ are different:
\begin{equation}
\square_{\rm MEP} = \square_{\rm pure} - \dfrac{1}{2 Z^2 \rho_{00}} \; .
\label{eq:diff_MEP_AAM_BnS}
\end{equation}
Despite the fact that $Z$ is a function of only the effective properties, the effective state $\rho$ in the present case, we were not able to obtain an analytical expression $Z=Z(\rho)$ -- the imposition in Eq.\eqref{eq:constraints_MEP_BnS}, or the equivalent set of equations in \eqref{eq:constraints_MEP}, lead to transcendental equations. We then resort to numerical evaluations in order to obtain the assigned state by the Maximum Entropy Principle for the situation described by the Blurred and Saturated detector.

\begin{figure}
\includegraphics[width=\linewidth]{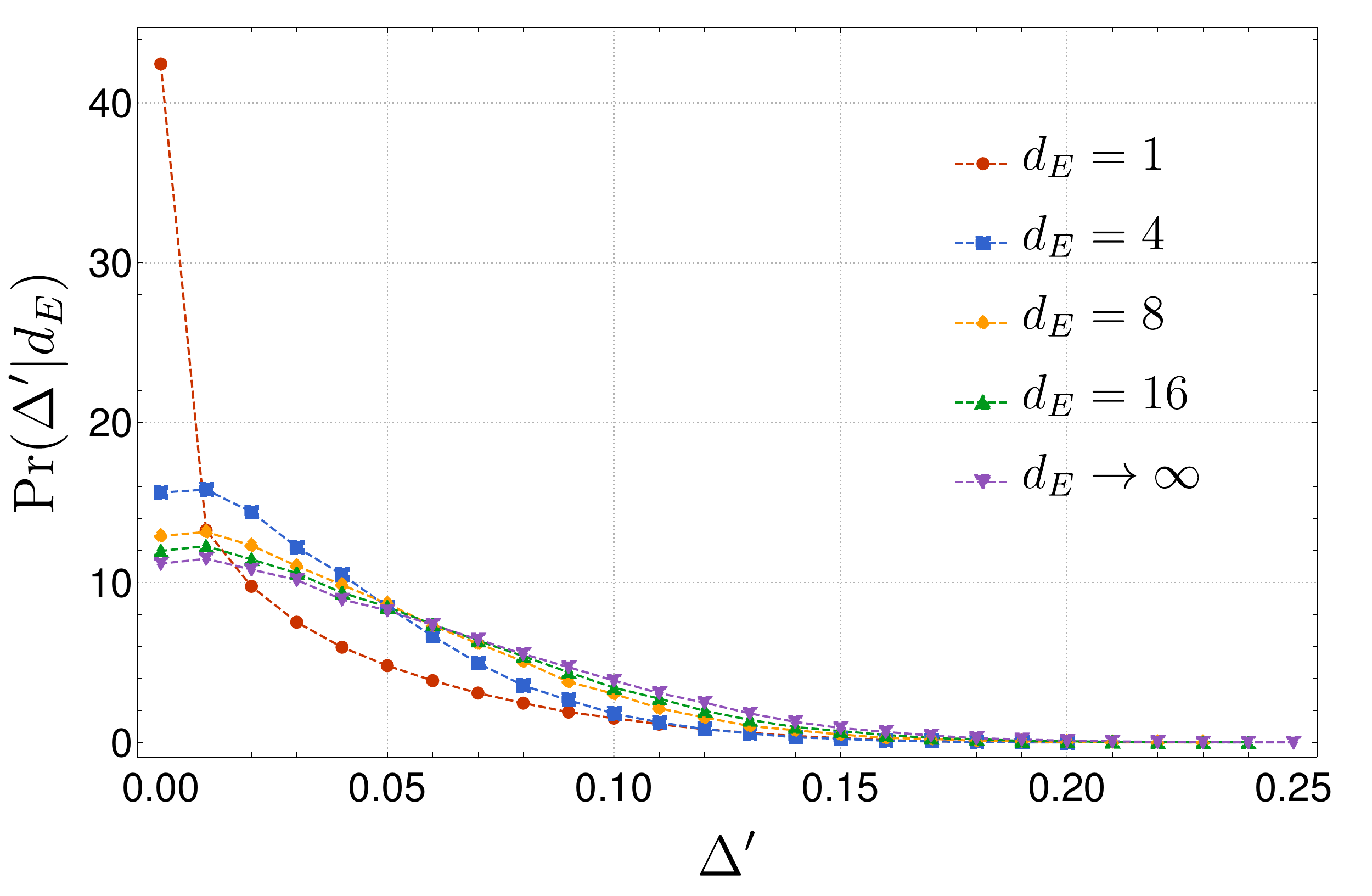}
\caption{\textbf{Comparison between the MEP and AAM assignments: $\Lambda_\text{BnS}$.} 
    The curves represent  the probability density $\pr(\Delta^\prime|d_E)$ of obtaining a distance $\Delta^\prime$ between the MEP and AAM assignments, given a fixed environment dimension $d_E$, when sampling effective states uniformly from the Bloch sphere.  The distance between the MEP and the pure-prior AAM assignment  corresponds to $d_E=1$ (red circles). The distance between the MEP and the mixed-prior AAM assignments are labeled according to the corresponding environmental 	dimension: $d_E=4,8,16,$ and $\infty$.}
\label{fig:Histogram}
\end{figure}

Comparisons between MEP and  AAM states, for both pure and mixed priors, are shown in Fig.~\ref{fig:Histogram}. It is interesting to observe that the probability of  small distances  ($\Delta^\prime\lesssim 0.02$) between the assigned state by the MEP and the AAM is larger for smaller environment dimension. Further numerical analyses are shown in Appendix~\ref{app:numerics}.

\subsection{SU(2) preserving coarse-graining map}
\label{subsec:ibrahim}

In the previous section we saw that the MEP and the AAM, despite similar guiding principles, may lead to different assigned states. However, the size of the underlying system in the Blurred and Saturated detector scenario was small, a two spin-1/2 system. One may wonder whether such differences still prevail when the dimension of the fine-grained system is much larger than the coarse-grained description.

To address such issue, in this section we exploit the coarse-graining map $\Lambda_J:\mathcal{L}(\mathcal{H}_D) \rightarrow \mathcal{L}(\mathcal{H}_2)$  introduced in Ref.~\cite{ibrahim}. The scenario described by $\Lambda_J$  is that of a system (that can be a single system, or a collection of subsystems) with total angular momentum $J$, which is then perceived as an effective single spin-1/2 system. The coarse-graining map $\Lambda_J$ is defined as
\begin{equation}
	\Lambda_J[\psi]=
    	\dfrac{1}{2}\bigg(\mathds{1}+\dfrac{1}{j}\sum_{i\in\{x,y,z\}}\tr({\psi}J_i) \,\sigma_i\bigg),
	\label{eq:IbraCG}
\end{equation}
where $J_i$ are the angular momentum components, i.e., the generators of SU(2) rotations around the $x,y,z$ axes in the $D$-dimensional Hilbert space $\mathcal{H}_D$, and 
$\sigma_i$ are the usual Pauli matrices. 
In the above expression, $j$ is the largest eigenvalue of $J_i$, thus fixing  $D=2j+1$.

An important property of this coarse-graining channel is that it maps 
a rotated microscopic state in $\mathcal{L}(\mathcal{H}_D)$ 
to a rotated qubit in $\mathcal{L}(\mathcal{H}_2)$ around the same axis and by the same angle, i.e., $\Lambda_J$ preserves the $SU(2)$ structure. Mathematically, this invariance is expressed by:
\begin{align}
	\Lambda_J[R_{\hat{n}}(\theta){\psi}R_{\hat{n}}(\theta)^\dagger] =
	R{'}_{\hat{n}}(\theta){\Lambda_J[\psi]}R{'}_{\hat{n}}(\theta)^\dagger \; ,
	\label{eq:rotinv}
\end{align}
where 
$R_{\hat{n}}(\theta)\in\mathcal{L}(\mathcal{H}_D)$ and 
$R_{\hat{n}}'(\theta)\in\mathcal{L}(\mathcal{H}_2)$ 
are rotations by an angle $\theta$ around a given axis $\hat{n}$:
\begin{align}
	R_{\hat{n}}(\theta)    & = \exp({-i{\theta}J_{\hat{n}}})           \; , \\
	R{'}_{\hat{n}}(\theta) & = \exp({-i{\theta}\sigma_{\hat{n}}/2})    \; ,
	\label{eq:rots}
\end{align} 
with $J_{\hat{n}}=\vec{J}\cdot\hat{n}$, $\sigma_{\hat{n}}=\vec{\sigma}\cdot\hat{n}$, 
and we are setting $\hbar=1$.  

As before we assume that an effective state $\rho$ is prepared, and we want to determine which state we should assign to the underlying system. Microscopically, the set of fine-grained states that abide by the constraints is written as:
\[
\Omega_{\Lambda_J}(\rho)=\{\psi \in \mc{L}(\mc{H_D}) \;|\; \Lambda_J(\psi)=\rho\}.
\]

\subsubsection{AAM assignment}

Before we dive in the evaluation of the AAM assignments (pure and mixed priors), two general observations are in place. First, note that writing $\rho$ in its Bloch representation,
\[\rho=\frac{1}{2}\left(\idty + \vec{r}.\vec{\sigma}\right),\]
with $\vec{r}.\vec{\sigma}\equiv r_x\sigma_x+r_y\sigma_y+r_z\sigma_z$, the coarse-graining constraint $\Lambda_J[\psi]=\rho$ can be expressed as
\begin{align}
\dfrac{1}{j}\tr({\psi}\vec{J}) = \vec{r}.
\label{eq:trchij}
\end{align}

Second, notice that if $\psi \in \Omega_{\Lambda_J}(\rho)$, then  $R_{\hat{r}}(\theta) {\psi} R_{\hat{r}}(\theta)^\dagger$ also belongs to $\Omega_{\Lambda_J}(\rho)$. This follows from the rotational invariance, Eq.\eqref{eq:rotinv}, when applied to a rotation around the axis defined by the  effective state Bloch vector:
\begin{equation}
	\Lambda_J[ R_{\hat{r}}(\theta) {\psi} R_{\hat{r}}(\theta)^\dagger] = R{'}_{\hat{r}}(\theta)\rho R{'}_{\hat{r}}(\theta)^\dagger=\rho.
	\label{eq:rotsymm}    
\end{equation}
Even though $R_{\hat{r}}$ is not a symmetry of $\Lambda_J$, as defined in \eqref{eq:def_symm} (because it is $\psi$-dependent), the above property simplifies the calculation of average assignments. 

This property can be applied in the general form of the AAM assignment, Eq.~(\ref{eq:ALambda}), by making  the change of variables $\psi \to R_{\hat{r}}(\theta) {\psi} R_{\hat{r}}(\theta)^\dagger$, to arrive at
\begin{equation}
	\mathcal{A}_{\Lambda_J}[\rho] = 
	\int d\mu_{\psi} \delta (\Lambda_J[\psi]-\rho) \, R_{\hat{r}}(\theta) {\psi} R_{\hat{r}}(\theta)^\dagger \; .
\end{equation}
As the assignment in this case, $\mathcal{A}_{\Lambda_J}[\rho]$, does not depend on $\theta$, we can average over the uniform distribution of rotations around the $\vec{r}$ axis, i.e., we can evaluate $\wick{\c1 R_{\hat{r}}(\theta) \psi \c1 R_{\hat{r}}(\theta)^\dagger}$ by integrating over $\theta\in[0,2\pi]$. Although we will not explicitly evaluate such an average, by expressing  $\psi$ in the common eigenbasis of $J_{\hat r}$ and $J^2$, which we denote simply by $\{\ket{m_{\hat{r}}}\}$ -- with $-j \le m \le j$ and where the sub-index $\hat{r}$ reminds the basis dependence on the Bloch vector of the effective state--,  it is immediate to see that
\begin{align}
	\wick{\c1 R_{\hat{r}}(\theta) \psi \c1 R_{\hat{r}}(\theta)^\dagger} &= \sum_{m,n}\<m_{\hat{r}}|\psi|n_{\hat{r}}\>\overline{e^{-\ii \theta(m-n)}}^{\mu_\theta}\ket{m_{\hat{r}}}\!\bra{n_{\hat{r}}}\nonumber\\
	&=\sum_m \<m_{\hat{r}}|\psi|m_{\hat{r}}\>\proj{m_{\hat{r}}}\;.\nonumber
\end{align}
In plain text, just using the invariance property, we determined that the AAM will assign to the fine-grained system a state which is diagonal in the basis $\{\ket{m_{\hat{r}}}\}$. We thus only need to calculate 
$D=2j+1$ averages.

Putting these two observations together, we arrive to the conclusion that the AAM for the coarse-graining map $\Lambda_J$ can be seen as a function of the effective state $\rho$ Bloch's vector (due to the rotational symmetry), and that it can be written as:
\[
\mc{A}_{\Lambda_J}[\vec{r}]=\sum_m p_m(r)\proj{m_{\hat{r}}}\;,
\]
with $p_m(r)\ge0$ and $\sum_m p_m(r)=1$ depending only on the modulus $r=|\vec{r}|$ of the Bloch vector due to the rotational symmetry. In what follows, we obtain the $p_m$ coefficients as a function of $r$.

\paragraph{Explicit integration method – pure-state measure.}

The evaluation of the AAM assigned state now proceeds very closely to the previous calculations. For the present coarse-graining map, $\Lambda_J$, using the observations above, we write 
\begin{equation}
	\mathcal{A}_{\Lambda_J}[\vec{r}] \propto \int d\mu_{\psi} 
			 \delta\left(\frac{1}{j}\tr[\psi \vec{J}]-\vec{r}\right) \psi \; .
\end{equation}
where for the moment, $d\mu_{\psi}$, is taken as the uniform measure over pure states acting in $\mc{L}(\mc{H}_D)$.

As the AAM assigned state is diagonal in the common basis of $J_{\hat{r}}$ and $J^2$, i.e., the basis $\{\ket{m_{\hat{r}}}\}$ defined above, we employ this basis to make the correspondence $\ket{\psi}\rightarrow c=(c_1,c_2,\cdots,c_D)^T$,  and to write the  angular-momentum matrices $\vec{\mathds{J}}$. 
The integral above then becomes 
\begin{equation}
	\mathcal{A}_{\Lambda_J}[\vec{r}] \propto \int d(c^\dagger,c) 
	        \, \delta\left( c^\dagger c - E \right)  
			\, \delta\left(\frac{1}{j} c^\dagger \vec{\mathds{J}} c -\vec{r} \right) c \, c^\dagger 
			 \equiv \mathcal{I} \; .
\label{eq:ass-ibra}
\end{equation}
Given that the condition $\tr[\psi \vec{J}]=j \vec{r}$ does not directly imposes the normalization of $\psi$, we explicitly included it with the first $\delta$. In the above, we introduced the variable $E > 0$ for mathematical convenience; at the end of calculations it  will be set $E=1$ \cite{laksh08}.

As before, we momentarily take $E$ and $\vec{r}$ as variables. We then perform a Laplace transform, and its inverse, in the normalization condition
\[\
\delta(c^\dagger c -E) =\mc{L}^{-1}\left\{e^{-s\;  c^\dagger c}\right\}(E),
\]
and a Fourier transformation, and its inverse, in the angular momentum constraints:
\[
\delta\left(\frac{1}{j} c^\dagger \vec{\mathds{J}} c -\vec{r} \right)=\mc{F}^{-1}\left\{e^{-\ii\,  c^\dagger\, \vec{k}.\vec{\mathds{J}}\, c/j} \right\}(\vec{r}).
\]
In the above, $s$ and $\vec{k}=(k_x,k_y,k_z)$ are the transformations variables.

Exchanging the order of integration between the inverse transforms and the $d(c^\dagger,c)$, we obtain equations analogous to Eqs.~\eqref{eq:N_det_pure} and \eqref{eq:I_det_pure}:
\begin{equation}
 \tilde{\mathcal{I}}_{ij}[s,\vec{k}]   =
	 \int d(c^\dagger,c) \, e^{-c^\dagger A c} \, c_i^\ast c_j = 
	   \dfrac{\pi^D}{\det A} A^{-1}_{ji}  \; , 
\label{eq:I_ibra_pure}
\end{equation}
with the corresponding normalization
\begin{equation}
 \tilde{\mathcal{N}}[s,\vec{k}]   =
	 \int d(c^\dagger,c^\dagger) \, e^{-c^\dagger A c} = \dfrac{\pi^D}{\det A} \; .
\label{eq:N_ibra_pure}
\end{equation}
For the present case,  the matrix $A$ is given by
\begin{align}
A(s,\vec{k}) = s\mathds{1} + \ii \frac{k}{j} \mathds{J}_{\hat{k}}  \; ,
\label{eq:det_ibra}
\end{align}
with $\mathds{J}_{\hat{k}}=\vec{\mathds{J}} \cdot \hat{k}$.

The next step is to calculate the inverse transforms. 
We were able to obtain analytical results, but the corresponding expressions are too 
lengthy to be shown here. 
Some details are presented in Appendix \ref{app3}. Additionally, Figure~\ref{fig:pm_ibra_pur} shows the diagonal elements $p_m$ of the pure-prior average assignment $\mathcal{A}_\Lambda[\vec{r}]$ for several values of $j$.
\begin{figure}
	\includegraphics[scale=0.175]{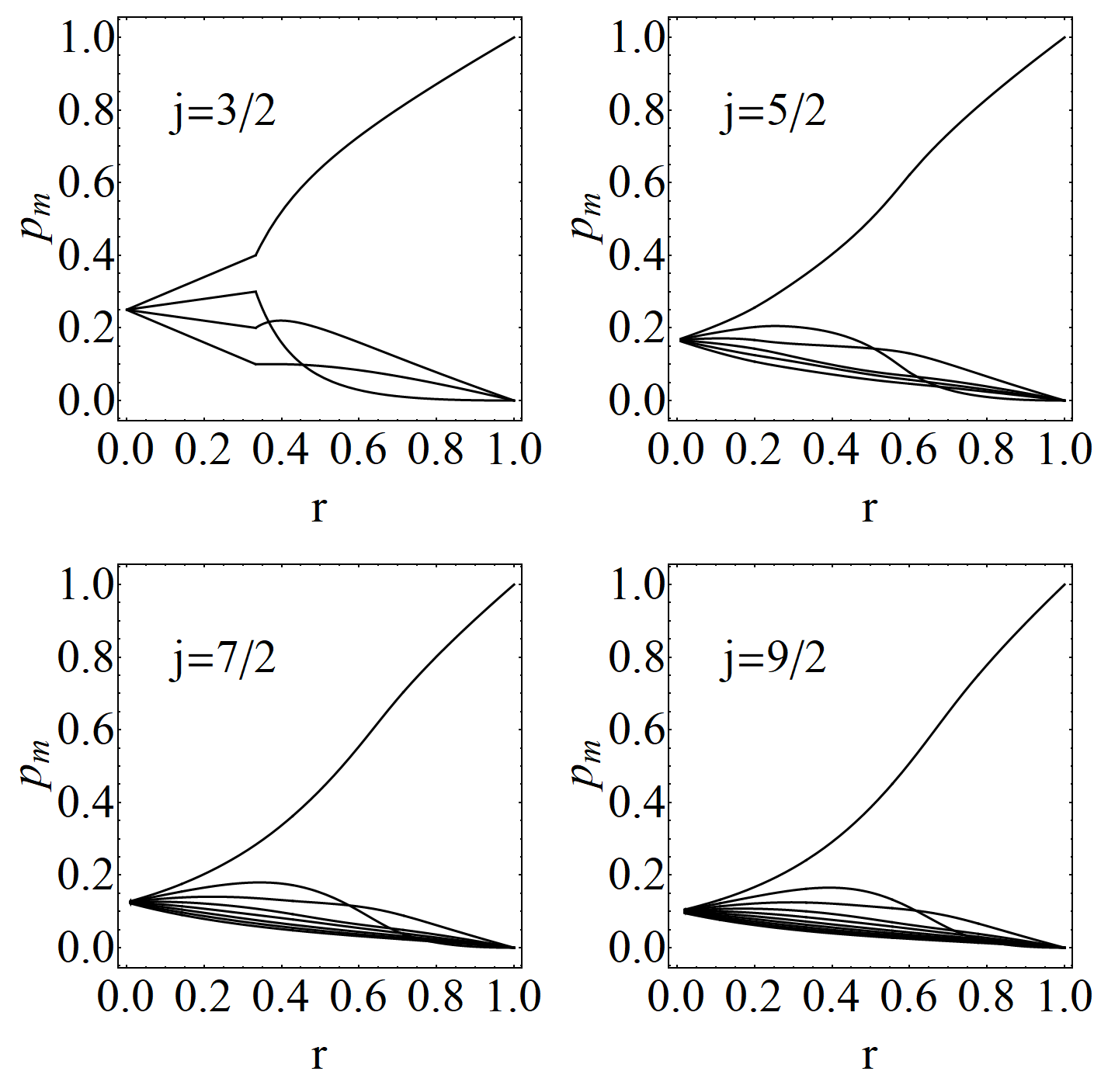}
	\caption{\textbf{Angular momentum distribution for the pure-prior average assignment.}
	Diagonal elements $p_m$ of $\mathcal{A}_{\Lambda_J}[\rho]$ are plotted as a 
	function of the parameter $r$. Each panel refers to a specific value of $j$: $j=3/2,5/2,9/2,7/2$ in clockwise direction.	In all panels, the curves $p_m(r)$ are ordered in decreasing order of $m$, 
	i.e, the topmost curve corresponds to $m=j$, and the lowest to $m=-j$.}
	\label{fig:pm_ibra_pur}
\end{figure}

\paragraph{Explicit integration method – mixed-state measure}
The evaluation of the AAM for the $\Lambda_J$ coarse-graining averaging  over mixed states proceeds in complete analogy to the case of the Blurred and Saturated detector (Sec.~\ref{sec:detector}). As in that case, here we express the results for mixed-states average, in terms of the pure-state case. The results are as follows:
\begin{equation}
	\tilde{\mathcal{I}}_{\rm mixed} = d_E \, \tilde{\mathcal{N}}^{d_E-1} \, \tilde{\mathcal{I}}[s,\vec{k}] \; , 
\label{eq:I_ibra_mixed}
\end{equation}
and 
\begin{equation}
	\tilde{\mathcal{N}}_{\rm mixed}   = \tilde{\mathcal{N}}^{d_E}[s,\vec{k}] \; , 
\label{eq:N_ibra_mixed}
\end{equation}
with $A$, $\tilde{\mathcal{N}}$ and $\tilde{\mathcal{I}}$ 
as in the previous section (pure case).

\begin{figure}[htp]
	\includegraphics[scale=0.17]{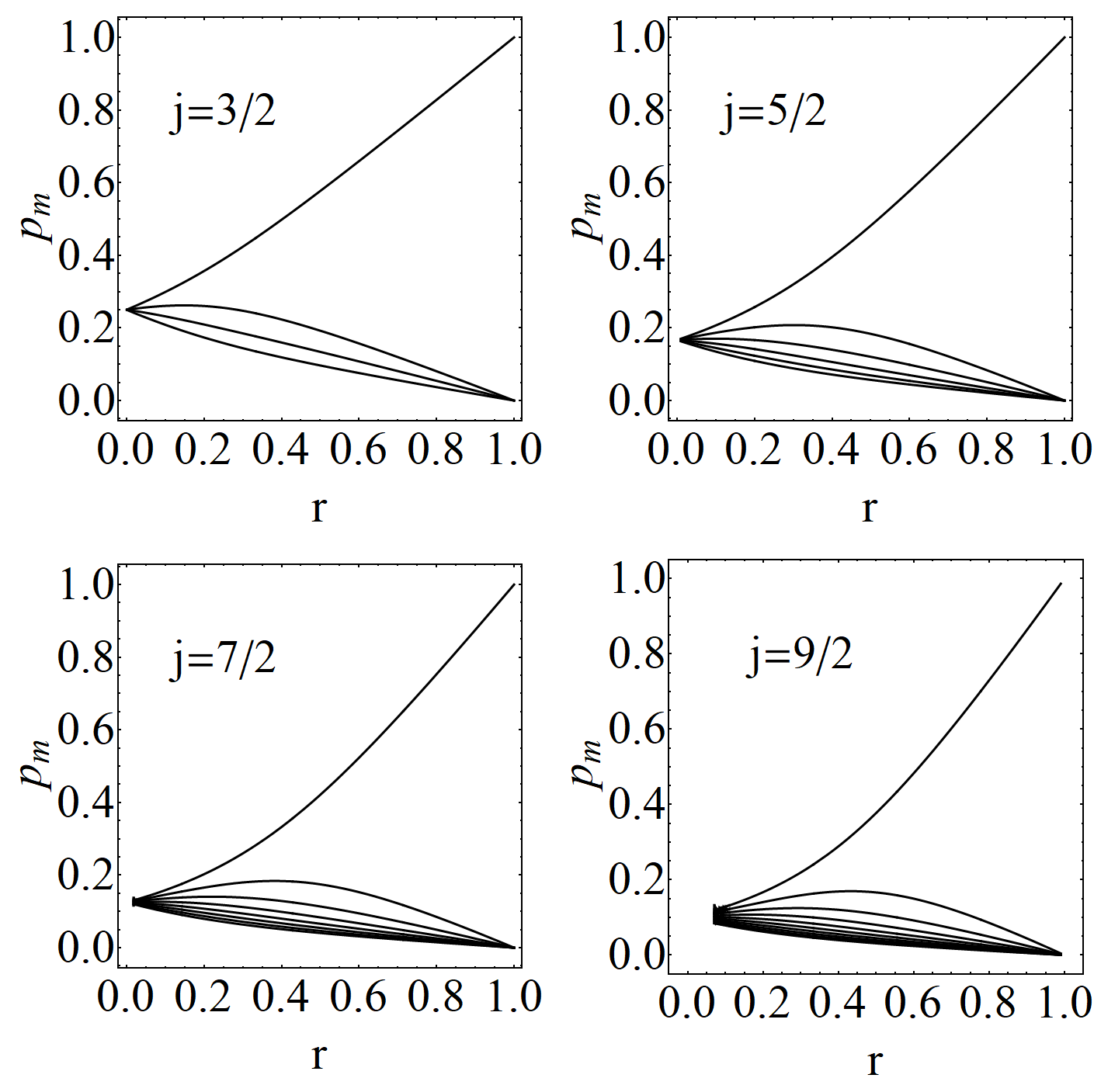}
	\caption{\textbf{Angular momentum distribution for a mixed-prior average assignment.}
	Diagonal elements $p_m$ of $\mathcal{A}_{\Lambda_J}[\rho]$ are plotted as a 
	function of the parameter $r$. 
	The curves in each panel are ordered in decreasing order of $m$, 
	i.e, the topmost curve corresponds to $m=j$, and the lowest to $m=-j$. 
	For each $j$-value we chose the dimension of the environment $d_E$ to be equal 
	to the system's $D=2j+1$.}
	\label{fig:pm_ibra_mix}
\end{figure}

We present some details about the inverse transformations and final results in Appendix~\ref{app3}.  
Diagonal elements $p_m$ of this assignment as a function of $r$ are plotted in 
Fig.~\ref{fig:pm_ibra_mix}. 
We chose to show the special case $d_E=D$ because that is the minimal dimension for which any mixed state that satisfies the constraints can be purified.

Some general features of the plots in Fig.~\ref{fig:pm_ibra_pur} and in Fig.~\ref{fig:pm_ibra_mix} can be readily understood. When $r=0$, i.e., the effective state is the maximally mixed state, no preferred direction is established. As such the fine-grained state, for both priors, is  the maximally mixed state in $\mc{L}(\mc{H}_D)$, and thus $p_m(0)=1/D$ for all $m$. On the other extreme, when the effective state is a pure state, $r=1$, then the set $\Omega_{\Lambda_J}(\rho)$ contains a single state, which is the state with maximum angular momentum in the same direction as the Bloch vector of $\rho$, namely $\proj{m_{\hat{r}}}$.

\subsubsection{MEP assignment}

Now we turn our attention to the determination of the state assigned by the Maximum Entropy Principle for the situation described by the coarse-graining map $\Lambda_J$. As for the AAM assignment, we assume knowledge of the effective state $\rho\in\mc{L}(\mc{H}_2)$. The MEP assignment for this case can then be expressed as the following optimization:
\begin{align}
    \psi_\text{MEP}^\rho =&\arg\max S(\psi)\\
                    & \text{s.t. }\;  \tr(\sigma_i \Lambda_J[\psi])=\tr(\sigma_i \rho)=r_i \;, \forall  i\in\{x,y,z\}.\nonumber
\end{align}
As explained in \ref{subsec:MEP}, under the above constraints, the MEP assignment is given by:
\begin{equation}\label{Eq:chimax}
\psi_{\rm MEP}^\rho= \frac{1}{Z}\exp{\left( -\sum_i \lambda_i\Lambda_J^*[\sigma_i] \right)}  \, ,
\end{equation}
 with the $\lambda_i$ related to the accessible values via: 
\begin{equation}
-\frac{\partial}{\partial \vec{\lambda}} \ln{Z} = \vec{r}  \, . 
\label{eq:const-ibra}
\end{equation}
To determine the action of the dual coarse-graining channel on the Pauli matrices, it is sufficient to notice that for any $\psi\in\mc{L}(\mc{H}_D)$, we have 
\begin{equation}
    \tr(\Lambda_J[\psi] \,\sigma_i )= \frac{1}{j}\tr(\psi \, J_i).
\end{equation}
It then follows that 
\begin{equation}
\Lambda^*[\sigma_i]=\frac{J_i}{j}  \; .
\end{equation}
That, in fact, is an alternative way of stating the rotational symmetry of $\Lambda_J$. Hence, the microscopic state that maximizes the entropy is given by 
\begin{equation}
\psi_{\rm MEP}^\rho = \frac{1}{Z} \exp{\left( -\frac{1}{j} \sum_i \lambda_i J_i      \right)}
              = \frac{1}{Z} \exp{\left( -\frac{1}{j} \lambda J_{\hat{\lambda}} \right)}  \; ,
\end{equation}
where  $J_{\hat \lambda} = \vec{J} \cdot \hat{\lambda}$, and $\lambda$ is the modulus of $\vec{\lambda}$

Imposing the constraint (\ref{eq:const-ibra}) we obtain at
\begin{equation}
-\frac{\partial}{\partial \vec{\lambda}} \ln{Z} = 
-\frac{\partial}{\partial      \lambda} \left( \ln{Z}\right) \, \hat{\lambda} = \vec{r}  \, . 
\end{equation}
From the above expression, we see that $\vec{\lambda}$ and $\vec{r}$ are parallel. Without loss of generality, we choose $\hat{\lambda} = -\hat{r}$ and arrive at 
\begin{equation}\label{Eq:MaxEntIbrahim}
\psi_{\rm MEP}^\rho = \frac{1}{Z} \exp{\left( \frac{1}{j} \lambda J_{\hat{r}} \right)}  \; .
\end{equation}
At the formal level this state is identical to the canonical ensemble of the Brillouin paramagnet
\cite{pathria}, with $\vec{\lambda}$ playing the role of the magnetic field  $\vec{B} = B \hat{r}$ (modulo some constant). Using this equivalence we can readily write the relation between $\lambda$ and $r$:
\begin{equation}
r = \frac{1}{j}\{ (j+\frac{1}{2})\coth{[(j+\frac{1}{2})\frac{\lambda}{j}]} 
                    - \frac{1}{2}\coth{\frac{\lambda}{2j}} \}= \mathcal{B}_j(\lambda) \; ,
\label{eq:brillouin}
\end{equation}
where $ \mathcal{B}_j$ is the so-called Brillouin function of $j$-th order.

\begin{figure}
	\centering
	\includegraphics[scale=0.175]{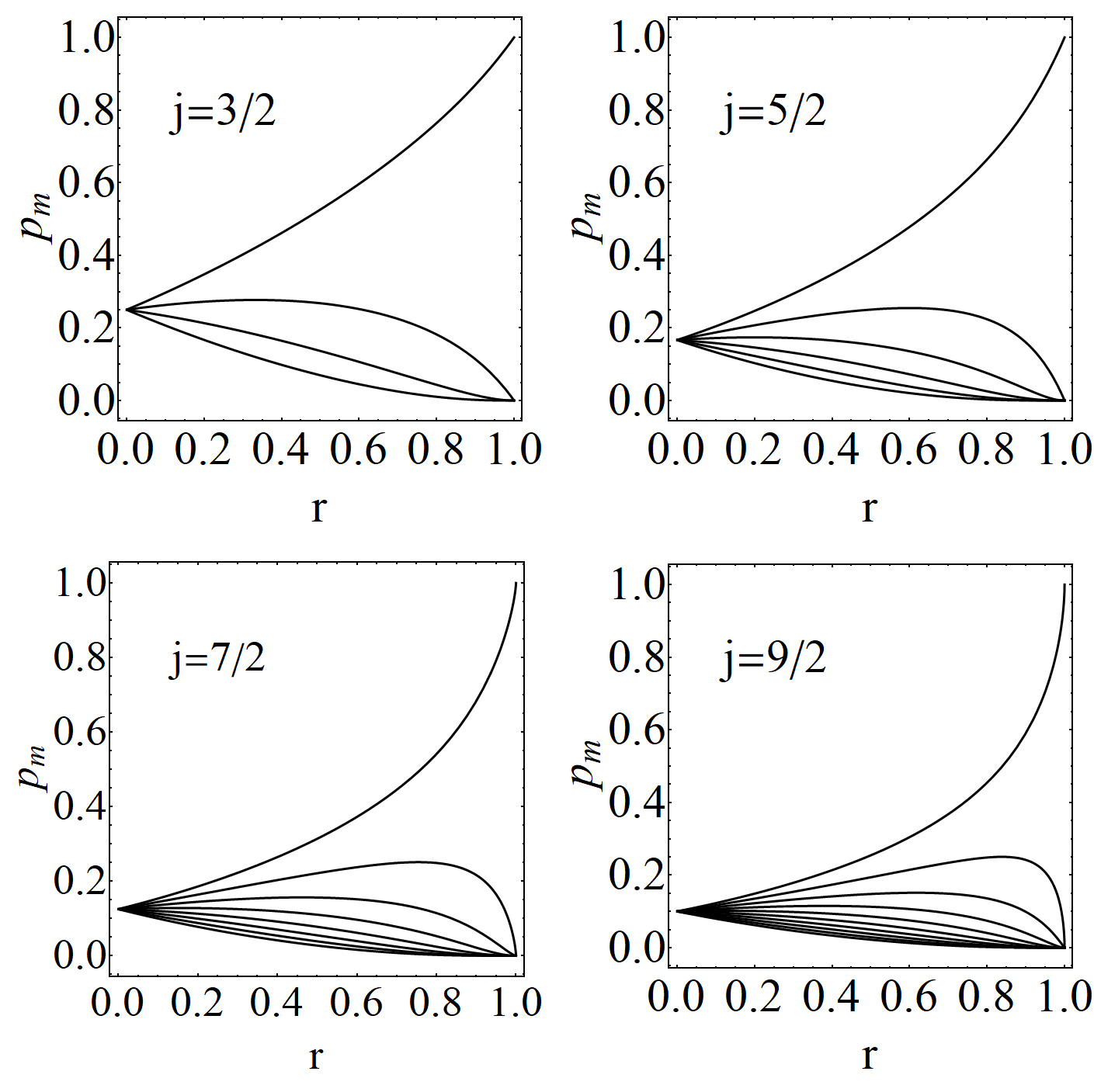}
	\caption{ \textbf{Angular momentum distribution for the MEP assignment.}
	Diagonal elements $p_m$ of $\psi_\text{MEP}^\rho$ are plotted as a 
	function of the parameter $r$. 
	Each one of the sub-figures correspond to a different $j$-value or system's dimension $D=2j+1$. 
	All curves $p_m(r)$ are ordered in decreasing order of $m$, i.e., the upper one corresponds to $m=j$, followed by the curves $m=j-1,\ldots,m=-j$.}
	\label{fig:ProbMaxEntropyStates}
\end{figure}

With the above we have all the elements to construct $\psi_\text{MEP}$ for the present case. Like for the AAM assignment, the state $\psi_{\rm MEP}$ is diagonal in the eigenbasis of $J_{\hat r}$ (and $J^2$), with matrix elements
\begin{equation}\label{eq:Ibrahim_MaxE_pm}
p_m(r) = \frac{1}{Z} \exp{\left( \lambda \frac{m}{j} \right)} \; ,
\end{equation}
with $m=-j,-j+1,\dots, j-1,j\/$, and 
\begin{equation}
Z = \frac{\sinh{[(j+1/2)\frac{\lambda}{j}]}}{\sinh{[\frac{\lambda}{2j}]}} \; .
\label{Eq:Zibraim}
\end{equation}
Here we did not explicitly show the $r$ dependence of $\lambda$ to avoid cluttering notation. Thus, the diagonal elements $p_m$ of the MEP assignment can be calculated from  Eqs.~(\ref{eq:brillouin}) and (\ref{eq:Ibrahim_MaxE_pm}). The results, for several values of $j$, are shown in Fig.~\ref{fig:ProbMaxEntropyStates}. 

\medskip

\begin{figure}
	\centering
	\includegraphics[width=\linewidth]{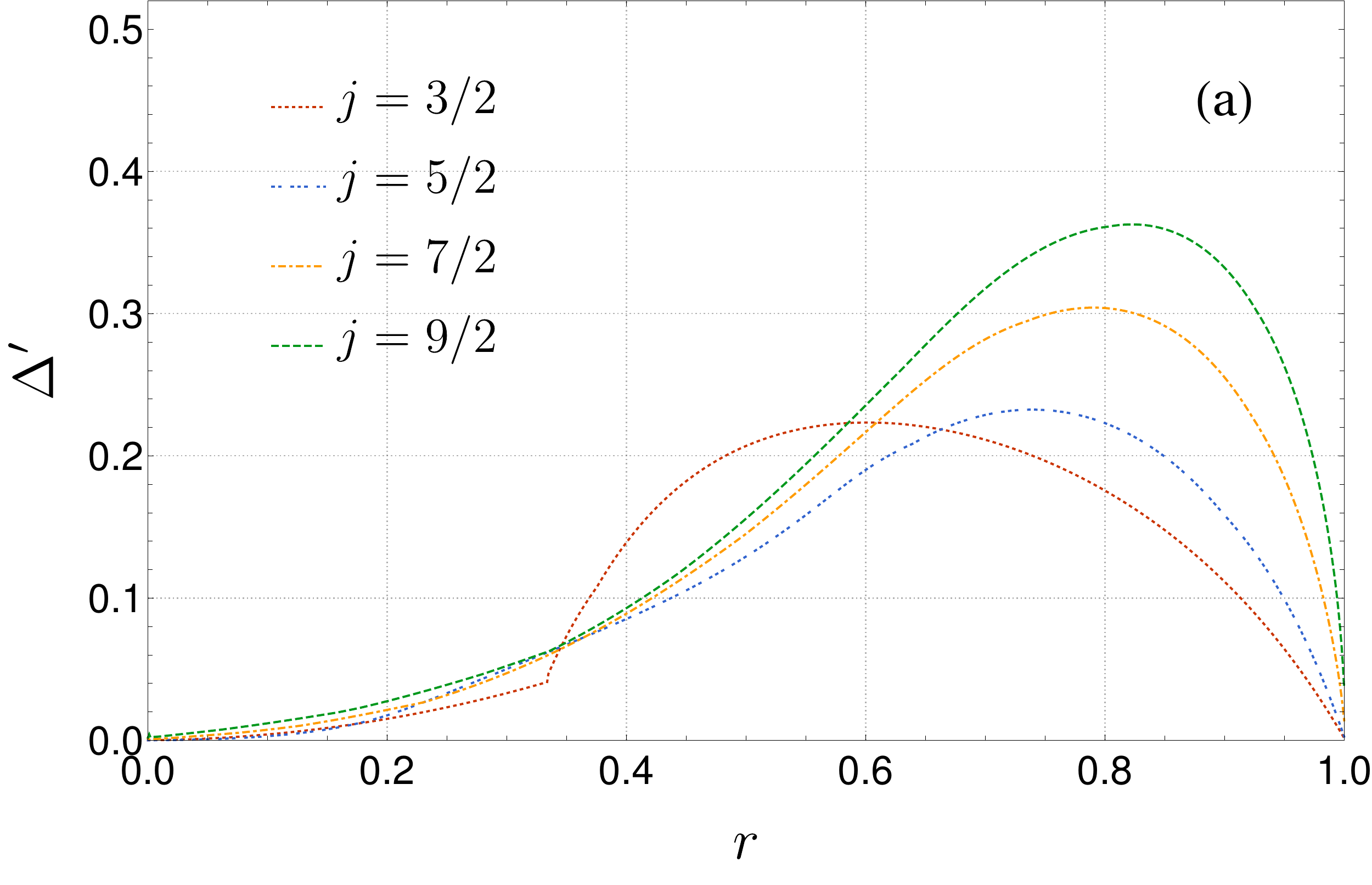}\\
	\includegraphics[width=\linewidth]{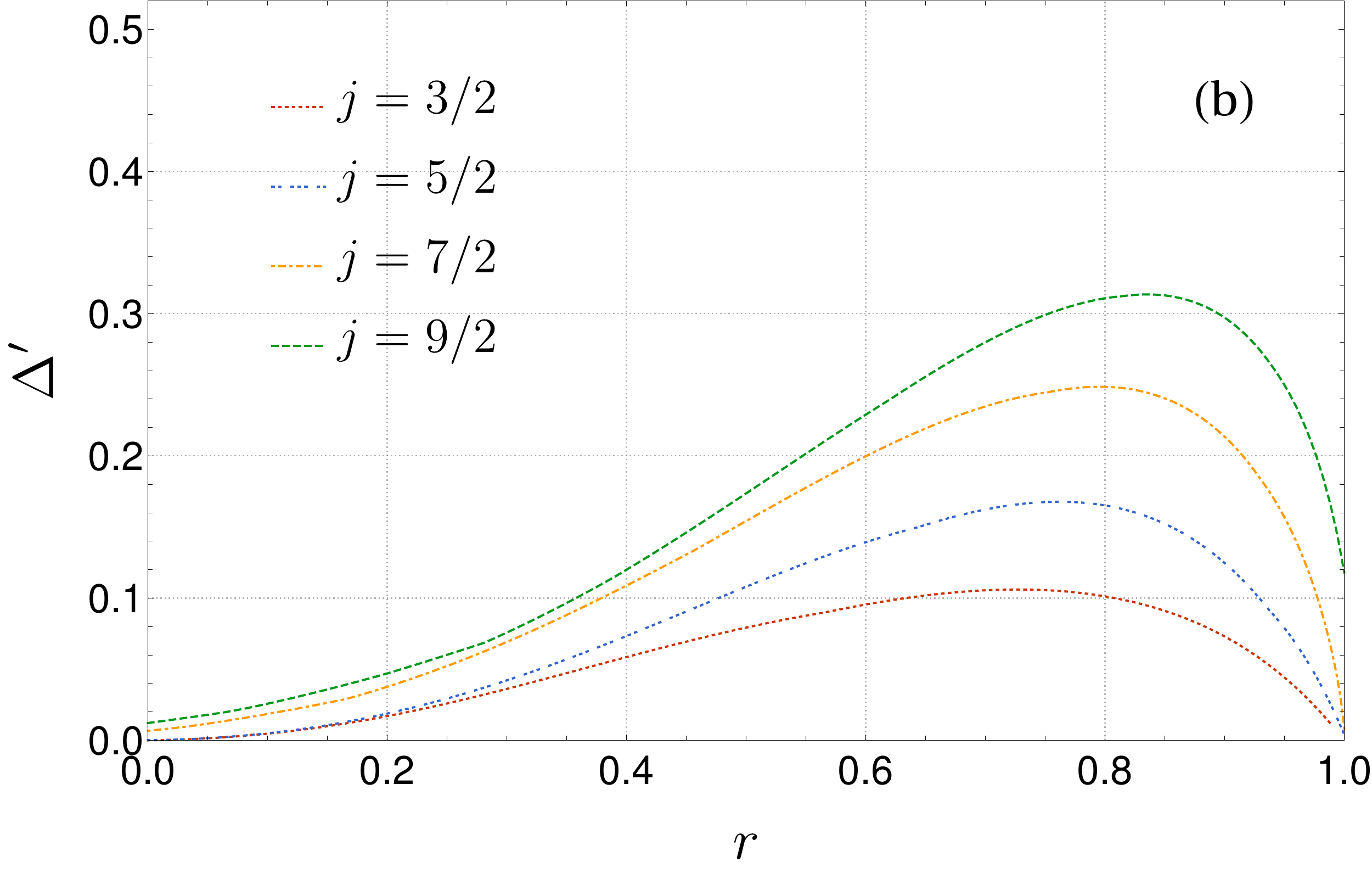}
	\caption{\textbf{Comparison between the MEP and AAM assignments: $\Lambda_J$.} 
    (a) Trace distance ($\Delta^\prime$) between the MEP and pure-prior AAM assignments 
	is plotted as a function of $r$. (b) Idem (a) but for mixed-prior AAM assignment. 
	We consider environmental dimension to be equal to the system's dimension, i.e., $d_E=2j+1$. 
	Several values of $j$ are illustrated.}
	\label{fig:comparisons}
\end{figure}

Overall, the MEP assignments are similar to the  AAM assignments -- with the same asymptotic behaviours at $r=0$ and $r=1$. A quantitative comparison is presented in Figure~\ref{fig:comparisons}, where we show the trace distance $\Delta^\prime$ between MEP and AAM assigned states as a function of $r$.  

\begin{figure}
	\includegraphics[width=\linewidth]{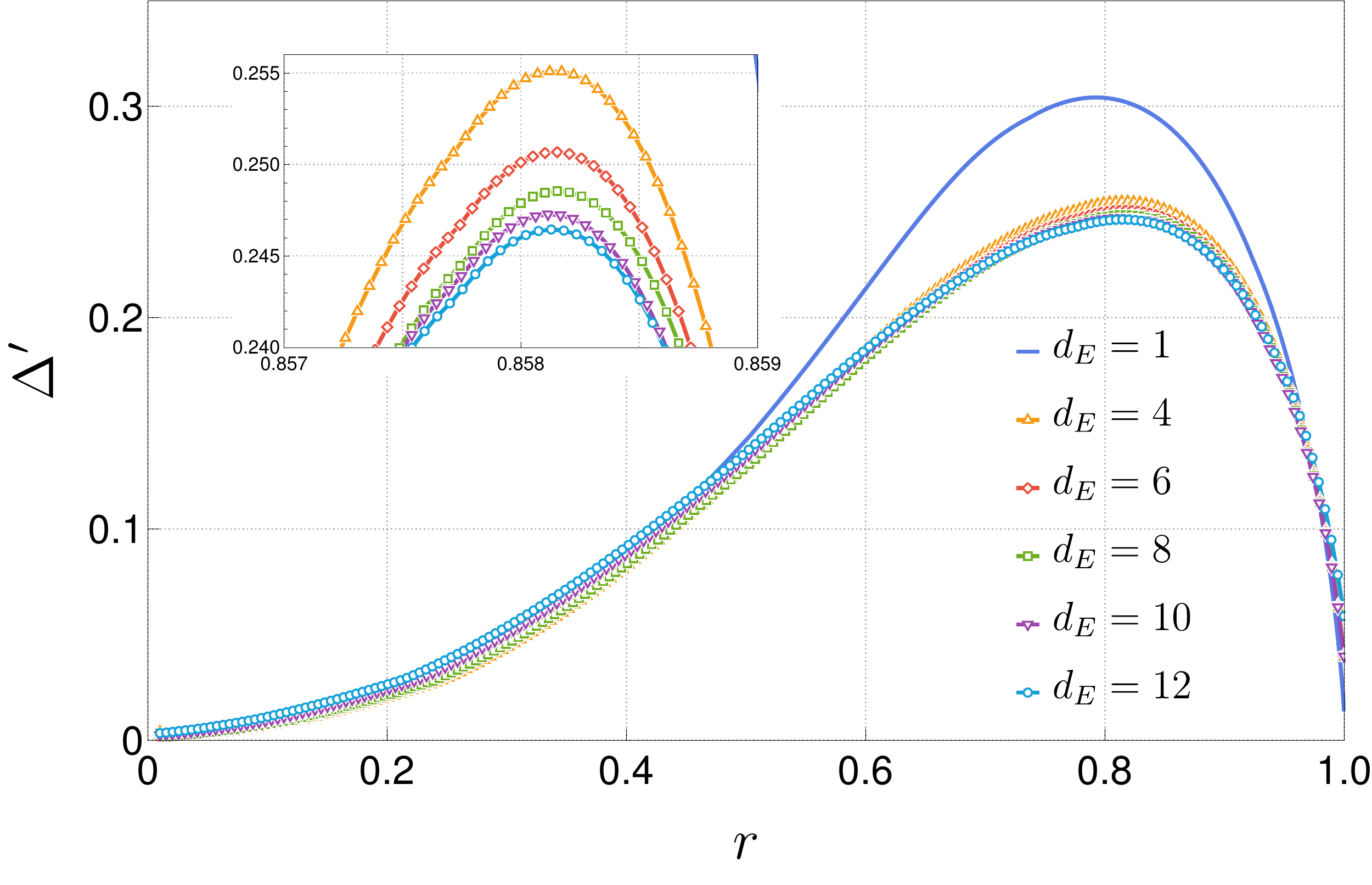}
	\caption{\textbf{Comparison between the MEP and AAM assignments.} 
    Trace distance ($\Delta^\prime$) between the MEP and AAM assignments for $j=7/2$ as a function of $r$. Several environmental dimensions are plotted. The inset shows the saturation of $\Delta^\prime$ as the environment's dimension grows.}
	\label{fig:varyingDE}
\end{figure}
Both cases, pure- or mixed-prior are similar, the trace distance being somewhat 
smaller for the mixed-case. However, as the dimension of microscopic system $j$ is increased, 
$\Delta^\prime$ keeps growing, with no sign of slowing-down. 

In Fig.\ref{fig:comparisons} the environmental dimension was chosen to be $d_E=2 j +1$, i.e., the same as the system dimension. In Fig.~\ref{fig:varyingDE} we show that the environment dimension plays no significant role in the distance between the MEP and AAM assignments -- a small decrease is observed, but the distance between the assignments is still appreciable, and it seems to saturate for large $d_E$. Further numerical analyses are shown in Appendix~\ref{app:numerics}.

\section{Application: Quantum Work}
\label{sec:Thermodynamics}

In the previous sections we have analyzed different schemes of quantum-state inference 
from coarse-grained information. This lead to different fine-grained states: the Average Assignment  Map (AAM) -- both pure and mixed priors--, and the Maximum Entropy Principle (MEP) assigned states. The aim of this section is to evaluate the observational consequences of these alternative assignments. In what follows, we exploit a thermodynamical process for which we assume that the effective state is given by the $SU(2)$ preserving 
coarse-graining map, and  inspect the impact of the different assignments in the amount of quantum work that can be extracted from the system.

Let us consider a microscopic quantum system composed of a particle, or system of particles, with total angular momentum $j$, which is thermally isolated. Furthermore, suppose that the underlying dynamics is given by the time-dependent Hamiltonian:
\begin{equation}
\label{Eq:Hsigmaquadrado}
H(t) = \gamma \, \cos(\omega t) J^{2}_{z},
\end{equation}
with $\gamma>0$, and $\omega$ an external driving frequency.

Now assume that we have access to this system only via the coarse-graining map $\Lambda_J$, defined in \eqref{eq:IbraCG}, with effective initial state given by:
\begin{equation}
    \rho_o=\frac{1}{2}\left(\idty+0.7\sigma_z\right).
    \label{eq:rho0}
\end{equation}
This initial state is chosen for two reasons. First, as we can see from Fig.\eqref{fig:comparisons}, the difference between the AAM and MEP assignments is considerable for all values of $j$. Second, the AAM and MEP procedures will lead to assigned states which are diagonal in the $J_z$ basis, and as such the assigned states will commute with the Hamiltonian. In this situation the quantum work is uncontroversially  defined by~\cite{binder2018QT,Colloquium2011quantumFluctuation,allahverdyan14,Talker2007}: 
\begin{eqnarray}\label{Eq:Workdef}
W=\tr\left(\overline{\psi}_0 \, (H_F-H_I)\right),
\end{eqnarray}
with $\overline{\psi}_0$ an initial assigned state (AAM-pure, AAM-mixed, or MEP) and
\begin{equation}
H_I\equiv H(0), \: \: H_F= U^{\dagger}_{\tau}H(\tau)U_\tau \; ,
\end{equation}
with $U_\tau$ the usual time-ordered evolution operator.

In such a scenario, the average quantum work reads
\begin{equation}\label{eq:work}
W = \gamma \, (\cos \omega \tau - 1) \, \tr \left( \overline{\psi}_0 J^{2}_{z} \right ) \; .
\end{equation}

\begin{figure}
 	\includegraphics[width=\linewidth]{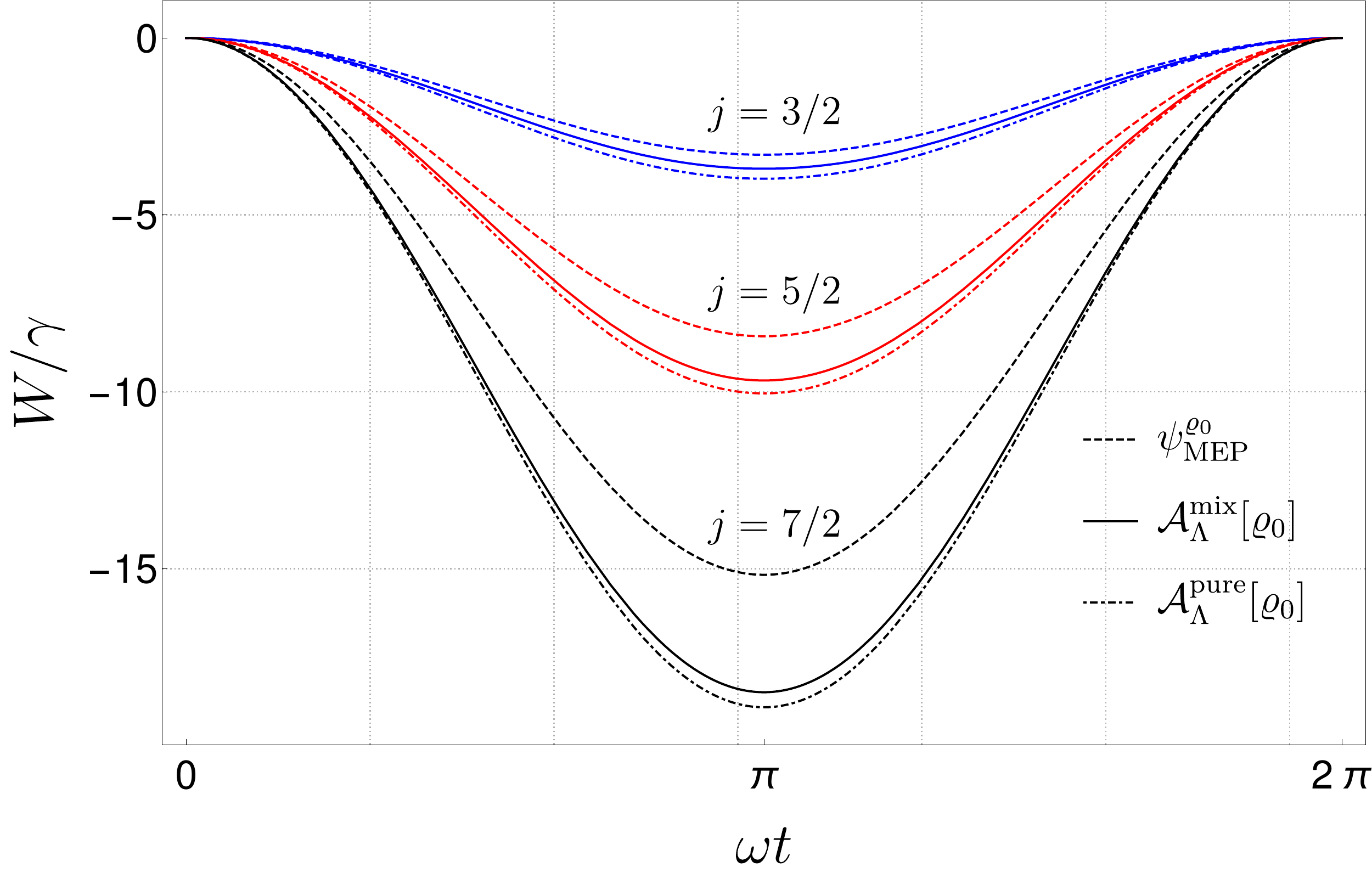}  
	\caption{\textbf{Average quantum work for different assignments.} The curves above shown the amount of work that can be extracted from the quantum system when only the effective  state is known, for different choices of assignments and different system sizes.	For each value of $j$ (3/2, 5/2, 7/2) three assignements are considered: 
	the MEP state (dashed line), the AAM state for a mixed prior ($d_E=2j+1$; solid line),
	and the AAM  state for a pure prior (dot-dashed line). In all the cases, the AAM state over pure states is the one that allows for more work extraction.}
	\label{fig:work}
\end{figure}

From the  expression above, it is clear that the difference between the assignments will translate into different values of average work.
The results for the quantum work are shown in Fig.~(\ref{fig:work}).

It is clear from Fig.~\eqref{fig:work} that using the MEP assignment always leads  to the least amount of work that can be extracted from the system. Moreover, among the AAM assignments, to use the average over pure states is always more beneficial, in terms of extracted work, than using the average over mixed states. This ordering can be understood by the knowledge of the physical scenario that each choice entails. While the MEP assignment is fixed given the constraints and the coarse-graining map, for the AAM assignments we still need to specify the measure (the prior) over the microscopic states that abide by the constraints. Choosing a pure state prior is consistent with an experimental setup where the microscopic system is fully isolated. A mixed state prior with a given environment dimension, on the other hand, assumes some lack of control about the microscopic system and some information about the environment. It is then reasonable to expect that from a situation where a pure state prior applies, one can extract more work from the quantum system, as it is the situation where more control about the system is present. Of course, using a prior which is not consistent with the experimental situation will lead to a mismatch between the amount of work predicted, and the one in fact extracted/produced.

\section{Conclusions}
\label{sec:Conclusions}

The maximum entropy principle (MEP) is ubiquitous, and rightly so as it is a cornerstone of thermodynamics and, more generally, of quantum statistical mechanics. Nevertheless, with the raise of complex quantum systems, new scenarios come into play. 

Here we analyzed how an assignment map, which shares some foundational ingredients with the MEP, namely the Average Assignment Map (AAM) behaves in various situations.  As expected, in the scenario where only local access to  a physical system is granted, the so called Open Quantum System picture, both assignments coincide.

However, in situations that go beyond the system-environment split, the two assignment strategies may differ. As we explicitly showed, in the case where two spin-1/2 particles, due to experimental constraints, are seen as an effective single spin-1/2, the two assignments differ -- see Eq.~\ref{eq:diff_MEP_AAM_BnS} and Fig.~\ref{fig:Histogram}.

As the underlying system is small, a two spin-1/2 particles, one may expect that such a difference would disappear for larger systems. With the coarse-graining map introduced in \ref{subsec:ibrahim}, we discredited this possibility. Even for increasing microscopic system dimensions, the difference between the MEP and the AAM assignments remains -- see Fig.~\ref{fig:comparisons}. As it can be seen, the difference in fact increases with the underlying system dimension in this situation.

Where does the difference come from?  Both assignments originate from the same idea: there are many microscopic states that abide by some few coarse-grained constraints. If no further constraints are present, the way we evaluate the expectation value of observable quantities in quantum mechanics suggests that an uniform average among all the states that satisfy the constraints is a sensible assignment. The MEP employs an entropic function to obtain such an assignment, with the idea that uniform mixtures maximize the entropy. 

The AAM, on the other hand, makes this averaging process explicit. In doing so it becomes evident that an uniform measure over the microscopic states must be chosen. As there is no unique uniform  measure among mixed states, the AAM allows for the introduction of  some prior knowledge about the experimental condition -- how isolated is the underlying system. If the microscopic system is very well isolated, one can take the average over pure states. If the environmental influence cannot be disregarded, an averaging over mixed states is more judicious. At this point the AAM connects to Bayesian inference methods~\cite{Caves2002,Blume2010}, and we show -- see for instance Figures \ref{fig:pofd} and \ref{fig:varyingDE} -- how the assignments differ for each prior knowledge situation.

Finally, when should one use the MEP or the AAM? We showed that depending on the access one has of the system, the assigned states may differ. Using the knowledge about how one access the system, and how the system interacts with its environment, if known, can only be beneficial. As different levels of knowledge about an experimental situation are often tied together with more efficient processes, we showed a thermodynamical process where the benefit of employing the AAM assignment is unmistakable (see Fig.~\ref{fig:work}). The gain of using the AAM assignment, instead of the MEP one, is specially clear in the case where one knows that the system is highly isolated. It should be remarked that choosing an assignment which theoretically would allow for a greater extraction of work, in an experiment scenario which is not befitting with such a choice, will only lead to a poor description and no gain.

\begin{acknowledgements}
We gladly acknowledge fruitful discussions with Adán Castillo Guerrero, Carlos Pineda, and David Dávalos. This work is supported in part by the National Council for Scientific and Technological Development,  CNPq Brazil (projects: Universal 406499/2021-7, and PCI-CBPF), and it is part of the Brazilian National Institute for Quantum Information. P.S.C.~acknowledges funding from the Air Force Office of Scientific Research under grant No. FA9550-19-1-0361.
\end{acknowledgements}

\appendix

\section{Symmetries of $\Lambda_\text{BnS}$}
\label{app1}

Let us rewrite the definition of symmetry (\ref{eq:def_symm}) as
\begin{equation}
 D \equiv	\Lambda[ U \chi U^\dagger] - \Lambda[\chi] = 0;
\end{equation}
 with matrix elements $D_{ij}$, where $i,j\in\{0,1\}$. As this identity must hold for all Hermitian matrices $\chi\in\mc{L}(\mc{H}_4)$, the $U$ dependent coefficients of $D_{ij}$ must all vanish.
From $D_{00}=0$, and assuming $\tr(\chi)=1$, we obtain $|U_{00}|=1$. Unitary then implies that the remaining elements of first column/row are null. Without loss of generality we can set $U_{00}=1$. Thus, up to now, we have
\begin{equation}
U=
    \begin{pmatrix}
        1 & 0 & 0         & 0 \\
        0 & \multicolumn{3}{c}{\multirow{3}{*}{\boxed{W}}} \\
        0 &  \multicolumn{3}{c}{}\\
        0 &  \multicolumn{3}{c}{} 
    \end{pmatrix}   \; ,
\end{equation}
where $W \in U(3)$, with its unitarity guaranteed  by the condition $D_{11}=0$.
From requesting $D_{12}=0$ we obtain the following equations
\begin{align}
    W_{11}+W_{21}+W_{31} & = 1  \; , \\
    W_{12}+W_{22}+W_{32} & = 1  \; , \\
    W_{13}+W_{23}+W_{33} & = 1  \; .
\end{align}
This means that $W$ possesses the normalized eigenvector $v_1=(1,1,1)/\sqrt{3}$ with unity eigenvalue, but it is otherwise arbitrary. Then, in a orthonormal basis $\{v_1,v_2,v_3\}$, the matrix $W$ has the structure
\begin{equation}
W=
    \begin{pmatrix}
        1 & 0           & 0  \\
        0 &\multicolumn{2}{c}{\multirow{2}{*}{\boxed{V}}}    \\
        0 &\multicolumn{2}{c}{}
    \end{pmatrix}  \; ,
\end{equation}
with $V$ an arbitrary unitary in $U(2)$ -- the arbitrariness of $V$ comes from  the freedom
in the  choice of $v_2$ and $v_3$.

The simplest way of calculating the average $\wick{\c1 U \chi \c1 U^\dagger}$ 
of Eq.~(\ref{eq:AvUXU_Det}) consists in choosing a parametrization of $U(2)$ \cite{zyczkowski94} and then integrating over the parameters.

\section{$\Lambda_\text{BnS}$ AAM assignment -- Details}
\label{app2}

Our starting point are the quantities $\tilde{\mathcal{N}}$ and $\tilde{\mathcal{I}}$, 
defined in Eqs.~\eqref{eq:N_det_pure} and~\eqref{eq:I_det_pure}, respectively.
These quantities, related to the average over pure states, are the basis also
for the mixed case calculations [see Eqs.~(\ref{eq:I_det_mix}) and (\ref{eq:N_det_mix}].

We begin by doing the inverse transforms in the normalization factor $\tilde{\mathcal{N}}$.
The determinant of the matrix $A$ reads:
\begin{equation}
  \det A =   s_1^2 (k_x^2 + k_y^2 + s_0 s_1) \equiv s_1^2 (\kappa^2 + s_0 s_1) \; ,
\end{equation}
the last equality defining $\kappa$. 
We first do the Laplace inverse-transform of $\tilde{\mathcal{N}}_{\text{mixed}}$ 
in the variable $s_0$:
\begin{equation}
  \g{L}^{-1}_{s_0} \left\{\dfrac{\pi ^{4 d_E}}{(\det A)^{d_E} }\right\}(\rho_{00}) = 
  \frac{\pi ^{4 d_E} \rho_{00}^{d_E-1} \exp{(-\kappa^2 \rho_{00}/s_1)}}
       {s_1^{3d_E} \Gamma (d_E)} \; .
\label{eq:N_lap_s0}
\end{equation}
Then, we Laplace inverse-transform the RHS above in $s_1$:
\begin{equation}
  \g{L}^{-1}_{s_1} \left\{ \ldots \right\}(\rho_{11}) = 
  \frac{\pi ^{4 d_E} \rho_{11}^{(3 d_E-1)/2} J_{3 d_E-1}(2\kappa \sqrt{\rho_{00} \rho_{11}})}
       {\kappa^{3d_E-1} \rho_{00}^{(d_E+1)/2}\Gamma (d_E)} \; ,
\label{eq:N_lap_s0s1}       
\end{equation}
where $J_{3 d_E-1}$ is a Bessel function.
Now we must do the inverse Fourier transforms on the last expression. 
Note that such expression only depends on the polar radius $\kappa$ in the plane $(k_x,k_y)$.
Using $r_\perp=\sqrt{x^2+y^2}$, we have
\begin{align}
 \mathcal{N}_{\text{mixed}} (\rho)
  & = \int dk_x \, dk_y e^{i (k_x x + k_y y)} (\ldots)                    \\
  & = 2 \pi  \int_0^\infty d\kappa \, \kappa \, J_0( \kappa r_\perp) (\ldots)   \\
  & = \frac{ 2^{4(1-d_E)} \pi^{4 d_E +1} (1-r^2)^{3 d_E-2} }
       {\Gamma (d_E) \Gamma (3 d_E-1) (1+z)^{2 d_E} }  \; ,
\end{align}
where we have made the substitutions $\rho_{00/11} \to (1 \pm z)/2$, used $x^2+y^2+z^2=r^2$, 
and $(\ldots)$ stands for the RHS of Eq.~(\ref{eq:N_lap_s0s1}).

In the following we apply the same succession of inverse transforms as above 
(in the same order) to $\tilde{\mathcal{I}}_{\text{mixed}}$.
We focus on the elements the six elements denoted with a square in Eq.~(\ref{eq:AvUXU_Det}), as those, for the mixed prior,  cannot be calculated using the symmetry method.
If we choose, say, $\mathcal{A}_{\Lambda,23}$ as a representative, we must 
inverse-transform $\tilde{\mathcal{I}}_{\text{mixed},23}$.
Combining Eqs.~(\ref{eq:N_det_pure}), (\ref{eq:I_det_pure}), and (\ref{eq:I_det_mix}),
we obtain:
\begin{align}
  \tilde{\mathcal{I}}_{\text{mixed},23} &  = 
  d_E \left( \dfrac{\pi^4}{\det A} \right)^{d_E} A^{-1}_{32} \\[7pt]
  & =   \dfrac{d_E \, \pi^{4 d_E}}{(\det A)^{d_E+1}} \, C_{23}  \; ,
\end{align}
where $C_{23}$ is the appropriate cofactor of $A$, which reads 
\begin{equation}
 C_{23}  = - \dfrac{\kappa^2 s_1}{3} \; .    
\end{equation}
The first Laplace inverse-transform, in the variable $s_0$, gives
\begin{equation}
  \g{L}^{-1}_{s_0} \left\{\tilde{\mathcal{I}}_{\text{mixed},23}\right\}(\rho_{00}) = 
 - \frac{\pi ^{4 d_E} \kappa^2 \rho_{00}^{d_E} \exp{(-\kappa^2 \rho_{00}/s_1)}}
      {3 \, \Gamma (d_E) s_1^{2(d_E+1)}} \; .
\label{eq:lap_s0}
\end{equation}
Transforming the expression above, now in $s_1$, produces
\begin{equation}
  \g{L}^{-1}_{s_1} \left\{ \ldots \right\}(\rho_{11}) = 
  -\frac{\pi ^{4 d_E} \rho_{11}^{(3 d_E+1)/2} J_{3 d_E+1}(2\kappa \sqrt{\rho_{00} \rho_{11}})}
       {3 \, \kappa^{3d_E-1} \rho_{00}^{(d_E+1)/2}\Gamma (d_E)} \; ,
\label{eq:lap_s0s1}       
\end{equation}
Now we calculate the inverse Fourier transform of the last expression: 
\begin{align}
 & \mathcal{I}_{\text{mixed},23}(\rho) = \nonumber \\[7pt]
 & \frac{ d_E 2^{3-4d_E} \pi^{4 d_E -1} (1-r^2)^{3 d_E-2}[3d_E(x^2+y^2)+z^2-1] }
       {3 \, \Gamma (d_E+1) \Gamma (3 d_E) (1+z)^{2d_E+1} }  \; .
\end{align}
Finally, forming the quotient $\mathcal{I}/\mathcal{N}$, we recover the result of
Eq.~(\ref{eq:ass_det_mix_pure}):
\begin{align}
 \mathcal{A}_{\Lambda,23}(\rho) & = 
  \frac{3 d_E \left( x^2+y^2 \right) + z^2 - 1}{6 (3 d_E-1) (z+1)} \\[7pt]
  & = \dfrac{d_E}{3 d_E-1} \dfrac{|\rho_{01}|^2}{  \rho_{00}} - \dfrac{\rho_{11}}{3(3 d_E-1)} \; .
\end{align}

\section{Distance between $\mc{A}_{\Lambda_\text{BnS}}^\text{mixed}$ and $\mc{A}_{\Lambda_\text{BnS}}^\text{pure}$}
\label{app2_trace}

The trace distance is just half of the trace norm of the difference of the matrices:
\begin{equation}
\Delta (\rho,\sigma) = {\frac {1}{2}}||\rho -\sigma ||_{1}
  ={\frac {1}{2}} \tr \left[{\sqrt {(\rho -\sigma )^{\dagger }(\rho -\sigma )}}\right]\; .
\end{equation}	
The purpose of the factor of two is to restrict the trace distance between two normalized 
density matrices to the range $[0, 1]$. 
Since density matrices are Hermitian,
\begin{equation}
\Delta (\rho,\sigma) = \frac {1}{2} \sum _{i}|\lambda _{i}| \; ,
\end{equation}	
where the $\lambda _{i}$  are eigenvalues of the Hermitian, but not necessarily positive, 
matrix $(\rho -\sigma )$.% \cite{wiki_trace}

The trace distance between average (over pure or mixed states) assignments can be calculated
from Eq.~(\ref{eq:ass_det_mix_pure}). 
The difference between assignments has the following structure:
\begin{equation}
\mathcal{A}_{\Lambda_\text{BnS}}^\text{mixed}[\rho] - \mathcal{A}_{\Lambda_\text{BnS}}^\text{pure}[\rho] = 
\begin{pmatrix}
		0 & 0     & 0    & 0   \\
		0 & 0     & \mu  & \mu \\
		0 & \mu   & 0    & \mu \\
		0 & \mu   & \mu  & 0   
	\end{pmatrix}\; .
\label{eq:det_diff}
\end{equation}	
From Eq.~(\ref{eq:ass_det_mix_pure}) we obtain
\begin{equation}
\mu = \dfrac{(d_E-1)(1-r^2)}{4(3 d_E-1)(1+z)}   \; .
\end{equation}	
The eigenvalues of the difference matrix (\ref{eq:det_diff}) are
$\{ 2 \, \mu, -\mu, -\mu, 0 \}$.
Then, the trace distance $\Delta$ is
\begin{equation}
\Delta= 2 \,\mu = \dfrac{(d_E-1)(1-r^2)}{2(3 d_E-1)(1+z)}   \; .
\end{equation}	
In order calculate the probability distribution of $\Delta$ we must 
compute the following integral over the Bloch ball:
\begin{equation}
\pr(\Delta|d_E) = \frac{3}{4 \, \pi} \int dx dy dz \, \delta \left( \Delta - 2 \mu(x,y,z;d_E) \right)  \; ,
\end{equation}	
which leads to Eq.~(\ref{eq:p_of_delta}).

\section{$\Lambda_J$ AAM assignments -- Details}
\label{app3}

In the case of the present coarse graining it is not possible to obtain general formulae 
of the average assignments for arbitrary $j$ and $d_E$. 
Each case must be handled separately.
In the following we show some details of the calculation of the inverse Fourier-
and Laplace-transforms of the quantities $\mathcal{\tilde{I}}$ and $\mathcal{\tilde{N}}$, 
whose quotient, $\mathcal{I}(\vec{r})/\mathcal{N}(\vec{r})$, determines the average assignment. 

We start by calculating the norm $\mathcal{N}(\vec{r})$.
The determinant (\ref{eq:det_ibra}) reads
\begin{equation}
\det A =  \prod_{m=-j}^j \left( s + \dfrac{i k m}{j}  \right) \; .
\end{equation}
Note that $\det A$ is even function of $k$.
We do first the inverse Fourier transform of $\tilde{\mathcal{N}}(s,\vec{k})$.
As $\det A$ only depends on $k=|\vec{k}|$, it is convenient to use spherical coordinates
in $\vec{k}$-space;
without loss of generality, we choose $\hat{z}=\hat{r}$.
Then, the Fourier transform is given by
\begin{equation}
\mathcal{N}_{\rm mixed}(s,\vec{r})  =  
     \dfrac{4 \, \pi^{D d_E + 1}}{r}  
     \int_0^\infty  dk \, \dfrac{k \, \sin (k \, r)}{(\det A)^{d_E}} \; . 
\end{equation}
Next, we Laplace transform:
\begin{equation}
\mathcal{N}_{\rm mixed}(\vec{r}) = 
\g{L}^{-1}_s \left\{\mathcal{N}_{\rm mixed}(s,\vec{r}) \right\}(E)|_{E=1} \; .
\end{equation}
By virtue of the SU(2) symmetry, the norm only depends on the modulus of $\vec{r}$,
for all $j$ and $d_E$.
We show one example, for $j=3/2$ and $d_E=2$:
\begin{align}
 \mathcal{N}_{\rm mixed}(s,r) & = 
    \frac{27 \pi^{10} e^{-3 r s}}{128 \, r s^6} \left[ 2 r s+e^{2 r s} (6 r s-3)+3\right] \; , \\
 \mathcal{N}_{\rm mixed}(r)   & =  
    \frac{9 \pi ^{10}}{5120r} 
        \left[ 3 (1-3 r)^5 \theta (1-3 r) + \right. \nonumber \\[7pt] 
                              & \mkern-48mu
        \left. 10 r (1-3 r)^4 \theta (1-3 r)+3 (r-1)^5 +30 (r-1)^4 r \right] \; .
\end{align}
As $j$ and/or $d_E$ are increased, the expressions above grow rapidly, though preserving
the structure: $\mathcal{N}_{\rm mixed}(r)$ is a sum of Heaviside theta functions
multiplied by polinomials of $r$; the theta functions increase in number with $j$, the polinomials
increase in degree with $j$ and $d_E$.

Now, let us turn to $\mathcal{\tilde{I}}_{\rm mixed}$.
Using Eqs.~(\ref{eq:N_ibra_pure}) and (\ref{eq:I_ibra_pure}) into (\ref{eq:I_ibra_mixed}), 
we arrive at
\begin{equation}
\mathcal{\tilde{I}}_{{\rm mixed},mm}(s,\vec{k}) = 
         d_E \, \pi^{d_E D} \dfrac{C_{mm}}{(\det A)^{d_E+1}} \; .
\end{equation}
We have observed that $C_{mm}$, the diagonal cofactors of $A$, 
do not depend on the azimuthal angle $\phi$; then, the inverse Fourier transform 
reads
\begin{align}
\mathcal{\tilde{I}}_{{\rm mixed},mm}(s,r) & = 
        2 d_E \, \pi^{d_E D+1} \int_0^\infty dk \dfrac{k^2}{(\det A)^{d_E+1}} \nonumber \\ 
                & \times \int_0^\pi d\theta \sin{\theta} \exp{(i k r \cos \theta)} \, C_{mm} \; .    
\end{align}
The last step consists in inverse-transforming Laplace ($s \to E$) the expression above, 
and setting $E=1$. 
The result, $\mathcal{I}_{{\rm mixed},mm}(r)$, has the same structure as the norm 
$\mathcal{N}_{\rm mixed}(r)$, but appreciably more complex.
Finally, dividing $\mathcal{I}_{{\rm mixed},mm}(r)$ by $\mathcal{N}_{\rm mixed}(r)$ 
we obtain the probabilities $p_m(r)$ plotted in the Figures~\ref{fig:pm_ibra_pur} and \ref{fig:pm_ibra_mix}.

\section{Further numerical analysis -- $\Lambda_\text{BnS}$ and $\Lambda_J$}
\label{app:numerics}

In the main text we showed that, for a fixed coarse-graining map $\Lambda$ and fixed effective state $\rho$,  the assignments when employing pure or mixed priors are in general different -- with the exception of the partial trace map. This conclusion can be visually apprehended, for instance, from Figs.~\ref{fig:pofd} and \ref{fig:comparisons}. For $\Lambda_\text{BnS}$ the analytical evaluation of the trace distance between $\mc{A}_{\Lambda_\text{BnS}}^\text{pure}(\rho)$ and $\mc{A}_{\Lambda_\text{BnS}}^\text{mixed}(\rho)$ is presented in Appendix\ref{app2_trace}.

In the above discussion it was, nevertheless, assumed that the effective state $\rho$ could be perfectly prepared, which of course it is not the case in a realistic scenario. Two points could then be raised:  \textit{i)} How close the error-free analytical assignment  $\mc{A}_\Lambda$ is from the one obtained when preparation errors are allowed? \textit{ii)} Does the difference between  the assigned states for pure and mixed priors prevail when errors in the preparation of  $\rho$ are allowed?

To address these points, we extended the definition of the set of states that abide by the constraints, as to allow for a preparation error $\epsilon$:
\begin{equation}
    \Omega^\epsilon_\Lambda(\rho)=\{\psi\in \mc{L}(\mc{H}_D)\;|\; ||\Lambda(\psi)-\rho||_1 \le \epsilon\}.
\end{equation}

We then proceed by the way of a simple Rejection Sampling algorithm: for fixed $\rho$, $\Lambda$ and $\epsilon$, we uniformly sampled a large number (to be specified later) of states in $\mc{L}(\mc{H}_D)$ and checked which ones belong to $\Omega^\epsilon_\Lambda(\rho)$. With the ones that are selected, we  estimate the assigned state with preparation error $\epsilon$ as the average state, $\mc{A}^\epsilon_\Lambda[\rho]= \overline{\Omega^\epsilon_\Lambda(\rho)}$.

Numerically, we generated a database of $10^6$ pure 2 qubits states, and $10^6$  pure 3 qubits states, sampled from their respective Haar measure.  These databases were used both for $\Lambda_\text{BnS}$ and $\Lambda_\text{J=3/2}$. With the two qubit sample we obtained $\mc{A}^{\epsilon,\text{pure}}_{\Lambda_\text{BnS}}$ and $\mc{A}^{\epsilon,\text{pure}}_{\Lambda_\text{J=3/2}}$ for selected effective states (see below). With the database of 3 qubits, after performing the partial trace on the third qubit, we estimated $\mc{A}^{\epsilon,\text{mixed}}_{\Lambda_\text{BnS}}$ and $\mc{A}^{\epsilon,\text{mixed}}_{\Lambda_\text{J=3/2}}$. In all  cases we set the preparation error as $\epsilon=0.025$.

\begin{figure}
   \includegraphics[width=\linewidth ]{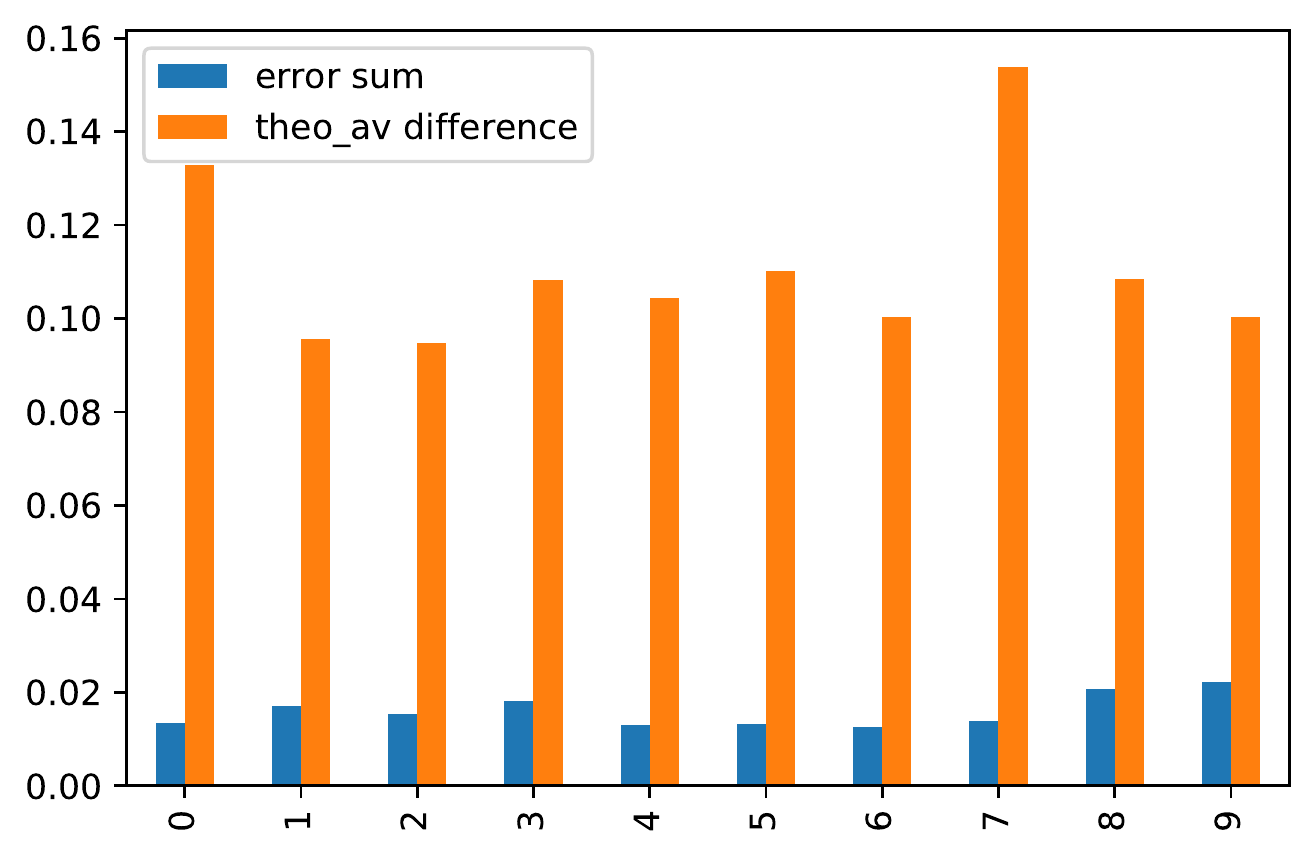}
	\caption{\textbf{Rejection Sampling analyses of preparation errors -- $\Lambda_\text{BnS}$.} The blue bars above (left bars for each state) denote the total error, $||\mc{A}^{\epsilon,\text{pure}}_{\Lambda}[\rho]-\mc{A}^{\text{pure}}_{\Lambda}[\rho]||_1+||\mc{A}^{\epsilon,\text{mixed}}_{\Lambda}[\rho]-\mc{A}^{\text{mixed}}_{\Lambda}[\rho]||_1$; while the orange bars quantify the distance between the error-free analytical results, $||\mc{A}^{\text{pure}}_{\Lambda}[\rho]-\mc{A}^{\text{mixed}}_{\Lambda}[\rho]||_1$. The horizontal axis represents the index of the effective state -- see table~\ref{tb:rhos}.}
	\label{fig:RS_BnS}
\end{figure}

\begin{figure}
    \includegraphics[width=\linewidth]{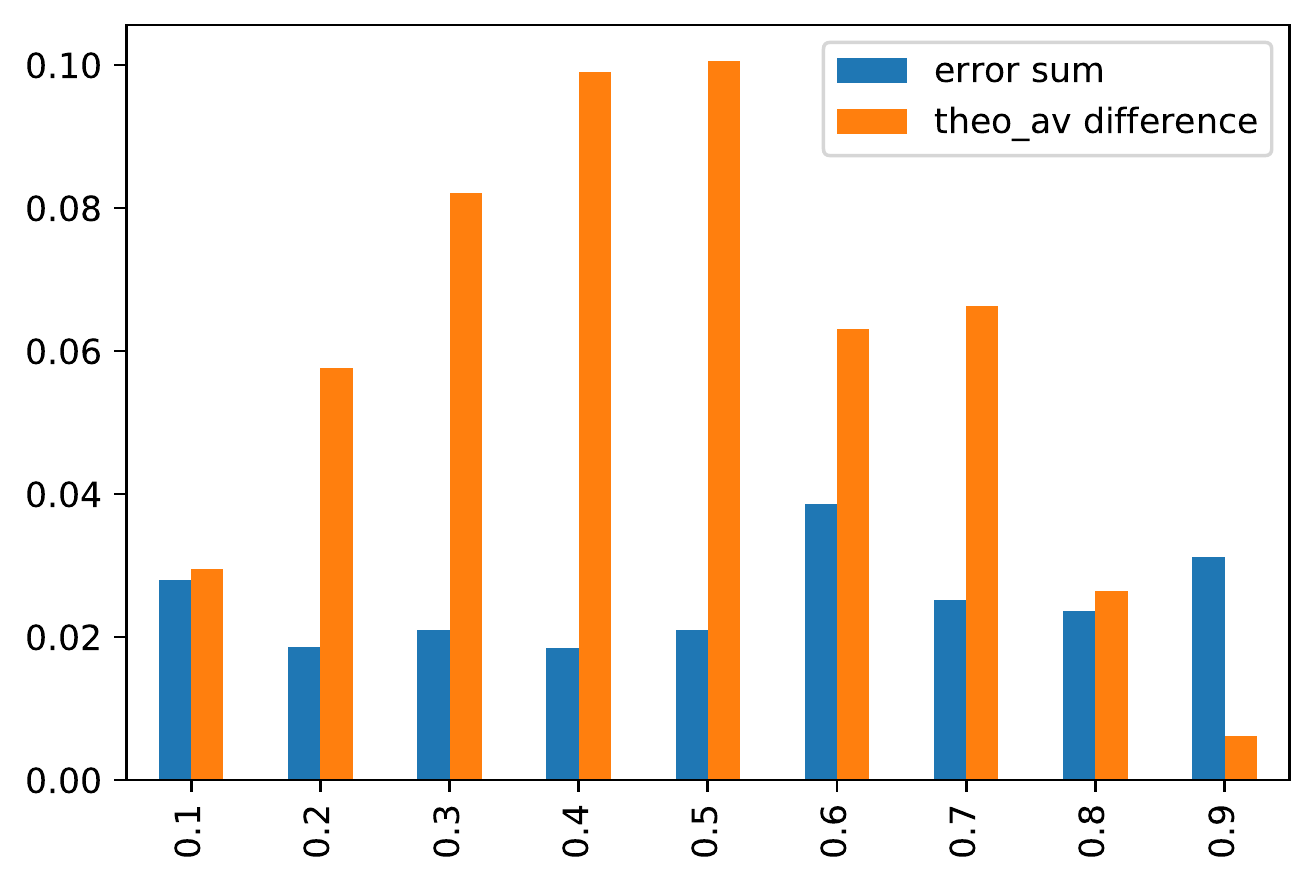}
	\caption{\textbf{Rejection Sampling analyses of preparation errors -- $\Lambda_\text{BnS}$. } The blue bars above (left bars for each state) denote the total error, $||\mc{A}^{\epsilon,\text{pure}}_{\Lambda}[\rho]-\mc{A}^{\text{pure}}_{\Lambda}[\rho]||_1+||\mc{A}^{\epsilon,\text{mixed}}_{\Lambda}[\rho]-\mc{A}^{\text{mixed}}_{\Lambda}[\rho]||_1$; while the orange bars quantify the distance between the error-free analytical results, $||\mc{A}^{\text{pure}}_{\Lambda}[\rho]-\mc{A}^{\text{mixed}}_{\Lambda}[\rho]||_1$. The horizontal axis represents the Bloch vector length $r$ of the effective state. }
	\label{fig:RS_J}
\end{figure}

Figures \ref{fig:RS_BnS} and \ref{fig:RS_J} summarize the obtained results. In both plots, the blue bars (left bar for each point in the horizontal axis)  quantify the total difference between the analytical expression and the ones obtained numerically, i.e., for each $\rho$ it equals $||\mc{A}^{\epsilon,\text{pure}}_{\Lambda}[\rho]-\mc{A}^{\text{pure}}_{\Lambda}[\rho]||_1+||\mc{A}^{\epsilon,\text{mixed}}_{\Lambda}[\rho]-\mc{A}^{\text{mixed}}_{\Lambda}[\rho]||_1$. The orange bars (right bar for each point in the horizontal axis) quantify the difference between the pure and mixed prior analytical states for each $\rho$, i.e., $||\mc{A}^{\text{pure}}_{\Lambda}[\rho]-\mc{A}^{\text{mixed}}_{\Lambda}[\rho]||_1$.

\begin{table}
\begin{tabular}{c | c c c } 
 \hline\hline
 index &$\tr(\rho \sigma_x)$&$\tr(\rho \sigma_y)$&$\tr(\rho \sigma_z)$\\
 \hline
0&-0.3061 &0.1269 & -0.6142 \\
1&0.0923& 0.1550& 0.0119\\
2&-0.0776&  0.1248&  0.03211\\
3&-0.2439&  0.0130& -0.1526\\
4&0.0749&  0.0032& -0.0502\\
5&-0.1384&  0.1779& -0.1613\\
6&-0.1082 &  -0.1748 &-0.0468\\
7&-0.1021&  0.0914& -0.5838\\
8&-0.1434& -0.1630& -0.1391\\
9&0.3696& -0.0652& -0.1729\\
\hline\hline
\end{tabular}
\caption{Effective states used in the Rejection Sampling algorithm to obtain the assigned states in the $\Lambda_\text{BnS}$ case when preparation error is allowed.}
\label{tb:rhos}
 \end{table}

This choice of presentation allows us to address both raised points. From the small values obtained for the blue bars we observe that the analytical error-free results are close to the ones when preparation errors are allowed. By comparing the height of the orange bars with the blue ones, we appreciate that the estimated average states with error are closer to their respective analytical result, than the distance between the analytical results for pure and mixed priors. In other words, the pure and mixed prior assigned states could be distinguished even if preparation errors are allowed.

As can be seen from the plot for $\Lambda_\text{BnS}$, Fig.~\ref{fig:RS_BnS}, for all randomly  selected states (see Table~\ref{tb:rhos}) the total error is always much smaller than the difference between the error-free analytical results. That confirms that when preparation errors are allowed, the assigned state does not change appreciably, and that pure and mixed prior assignments can be distinguished.

For  $\Lambda_\text{J}$, Fig.~\ref{fig:RS_J}, as expected, the difference between the error-free pure and mixed priors is small for almost pure effective states ($r\simeq1$), and highly mixed effective states ($r\simeq 0$). For such cases, the preparation error might not permit the distinction between pure and mixed prior assignments. This distinction is however possible for intermediate purity effective states.  In all the cases, given the small size of the blue bars, the assigned states when $\epsilon=0.025$ are close to the ones when $\epsilon=0$.

\bibliographystyle{unsrtnat}
\bibliography{ref}

\begin{thebibliography}{50}
\providecommand{\natexlab}[1]{#1}
\providecommand{\url}[1]{\texttt{#1}}
\expandafter\ifx\csname urlstyle\endcsname\relax
  \providecommand{\doi}[1]{doi: #1}\else
  \providecommand{\doi}{doi: \begingroup \urlstyle{rm}\Url}\fi

\bibitem[Jaynes(1957{\natexlab{a}})]{jaynes106}
E.~T. Jaynes.
\newblock Information theory and statistical mechanics.
\newblock \emph{Phys. Rev.}, 106:\penalty0 620--630, May 1957{\natexlab{a}}.
\newblock \doi{10.1103/PhysRev.106.620}.
\newblock URL \url{https://link.aps.org/doi/10.1103/PhysRev.106.620}.

\bibitem[Jaynes(1957{\natexlab{b}})]{jaynes108}
E.~T. Jaynes.
\newblock Information theory and statistical mechanics. ii.
\newblock \emph{Phys. Rev.}, 108:\penalty0 171--190, 1957{\natexlab{b}}.
\newblock \doi{10.1103/PhysRev.108.171}.
\newblock URL \url{https://link.aps.org/doi/10.1103/PhysRev.108.171}.

\bibitem[Banavar et~al.(2010)Banavar, Maritan, and Volkov]{banavar2010}
Jayanth~R Banavar, Amos Maritan, and Igor Volkov.
\newblock Applications of the principle of maximum entropy: from physics to
  ecology.
\newblock \emph{Journal of Physics: Condensed Matter}, 22\penalty0
  (6):\penalty0 063101, 2010.
\newblock \doi{10.1088/0953-8984/22/6/063101}.
\newblock URL \url{https://doi.org/10.1088/0953-8984/22/6/063101}.

\bibitem[Karmeshu(2003)]{karmeshu2003}
Jawaharlal Karmeshu.
\newblock \emph{Entropy measures, maximum entropy principle and emerging
  applications}, volume 119.
\newblock Springer Science \& Business Media, 2003.

\bibitem[Harte(2011)]{harte2011}
John Harte.
\newblock \emph{Maximum entropy and ecology: a theory of abundance,
  distribution, and energetics}.
\newblock OUP Oxford, 2011.

\bibitem[Boomsma et~al.(2014)Boomsma, Ferkinghoff-Borg, and
  Lindorff-Larsen]{boomsma2014}
Wouter Boomsma, Jesper Ferkinghoff-Borg, and Kresten Lindorff-Larsen.
\newblock Combining experiments and simulations using the maximum entropy
  principle.
\newblock \emph{PLoS computational biology}, 10\penalty0 (2):\penalty0
  e1003406, 2014.
\newblock \doi{10.1371/journal.pcbi.1003406}.
\newblock URL \url{https://doi.org/10.1371/journal.pcbi.1003406}.

\bibitem[{De Martino} and {De Martino}(2018)]{DeMartino2018}
Andrea {De Martino} and Daniele {De Martino}.
\newblock An introduction to the maximum entropy approach and its application
  to inference problems in biology.
\newblock \emph{Heliyon}, 4\penalty0 (4):\penalty0 e00596, 2018.
\newblock ISSN 2405-8440.
\newblock \doi{https://doi.org/10.1016/j.heliyon.2018.e00596}.
\newblock URL
  \url{https://www.sciencedirect.com/science/article/pii/S2405844018301695}.

\bibitem[Cesari et~al.(2018)Cesari, Reißer, and Bussi]{Cesari2018}
Andrea Cesari, Sabine Reißer, and Giovanni Bussi.
\newblock Using the maximum entropy principle to combine simulations and
  solution experiments.
\newblock \emph{Computation}, 6\penalty0 (1), 2018.
\newblock ISSN 2079-3197.
\newblock \doi{10.3390/computation6010015}.
\newblock URL \url{https://www.mdpi.com/2079-3197/6/1/15}.

\bibitem[Berger et~al.(1996)Berger, Della~Pietra, and Della~Pietra]{berger1996}
Adam Berger, Stephen~A Della~Pietra, and Vincent~J Della~Pietra.
\newblock A maximum entropy approach to natural language processing.
\newblock \emph{Computational linguistics}, 22\penalty0 (1):\penalty0 39--71,
  1996.
\newblock ISSN 0891-2017.

\bibitem[Bera and Park(2008)]{bera2008}
Anil~K. Bera and Sung~Y. Park.
\newblock Optimal portfolio diversification using the maximum entropy
  principle.
\newblock \emph{Econometric Reviews}, 27\penalty0 (4-6):\penalty0 484--512,
  2008.
\newblock \doi{10.1080/07474930801960394}.
\newblock URL \url{https://doi.org/10.1080/07474930801960394}.

\bibitem[Caticha and Golan(2014)]{caticha2014}
Ariel Caticha and Amos Golan.
\newblock An entropic framework for modeling economies.
\newblock \emph{Physica A: Statistical Mechanics and its Applications},
  408:\penalty0 149--163, 2014.
\newblock ISSN 0378-4371.
\newblock \doi{https://doi.org/10.1016/j.physa.2014.04.016}.
\newblock URL
  \url{https://www.sciencedirect.com/science/article/pii/S0378437114003276}.

\bibitem[Pathria and Beale(2011)]{pathria}
R.~K. Pathria and Paul~D. Beale.
\newblock \emph{Statistical Mechanics}.
\newblock Academic Press, 2011.

\bibitem[Bužek et~al.(1998)Bužek, Derka, Adam, and Knight]{buvzek1998}
V.~Bužek, R.~Derka, G.~Adam, and P.L. Knight.
\newblock Reconstruction of quantum states of spin systems: From quantum
  bayesian inference to quantum tomography.
\newblock \emph{Annals of Physics}, 266\penalty0 (2):\penalty0 454--496, 1998.
\newblock ISSN 0003-4916.
\newblock \doi{https://doi.org/10.1006/aphy.1998.5802}.
\newblock URL
  \url{https://www.sciencedirect.com/science/article/pii/S000349169895802X}.

\bibitem[Bu{\v{z}}ek(2004)]{buvzek2004}
Vladim{\'\i}r Bu{\v{z}}ek.
\newblock Quantum tomography from incomplete data via maxent principle.
\newblock In \emph{Quantum state estimation}, Lecture Notes in Physics,
  chapter~6, pages 189--234. Springer, Berlin, Heidelberg, 2004.
\newblock \doi{10.1007/978-3-540-44481-7_6}.
\newblock URL \url{https://doi.org/10.1007/978-3-540-44481-7_6}.

\bibitem[Caldeira and Leggett(1981)]{caldeira}
A.~O. Caldeira and A.~J. Leggett.
\newblock Influence of dissipation on quantum tunneling in macroscopic systems.
\newblock \emph{Phys. Rev. Lett.}, 46:\penalty0 211--214, Jan 1981.
\newblock \doi{10.1103/PhysRevLett.46.211}.
\newblock URL \url{https://link.aps.org/doi/10.1103/PhysRevLett.46.211}.

\bibitem[Breuer et~al.(2002)Breuer, Petruccione, et~al.]{breuer2002}
Heinz-Peter Breuer, Francesco Petruccione, et~al.
\newblock \emph{The theory of open quantum systems}.
\newblock Oxford University Press on Demand, 2002.

\bibitem[Mermin(1980)]{mermin1980}
N.~D. Mermin.
\newblock Quantum mechanics vs local realism near the classical limit: A bell
  inequality for spin $s$.
\newblock \emph{Phys. Rev. D}, 22:\penalty0 356--361, Jul 1980.
\newblock \doi{10.1103/PhysRevD.22.356}.
\newblock URL \url{https://link.aps.org/doi/10.1103/PhysRevD.22.356}.

\bibitem[Poulin(2005)]{poulin2005}
David Poulin.
\newblock Macroscopic observables.
\newblock \emph{Phys. Rev. A}, 71:\penalty0 022102, Feb 2005.
\newblock \doi{10.1103/PhysRevA.71.022102}.
\newblock URL \url{https://link.aps.org/doi/10.1103/PhysRevA.71.022102}.

\bibitem[Kofler and Brukner(2008)]{caslavLG}
Johannes Kofler and {\ifmmode \check{C}\else \v{C}\fi{}}aslav Brukner.
\newblock Conditions for quantum violation of macroscopic realism.
\newblock \emph{Phys. Rev. Lett.}, 101:\penalty0 090403, Aug 2008.
\newblock \doi{10.1103/PhysRevLett.101.090403}.
\newblock URL \url{https://link.aps.org/doi/10.1103/PhysRevLett.101.090403}.

\bibitem[Raeisi et~al.(2011)Raeisi, Sekatski, and Simon]{Raeisi2011}
Sadegh Raeisi, Pavel Sekatski, and Christoph Simon.
\newblock Coarse graining makes it hard to see micro-macro entanglement.
\newblock \emph{Phys. Rev. Lett.}, 107:\penalty0 250401, Dec 2011.
\newblock \doi{10.1103/PhysRevLett.107.250401}.
\newblock URL \url{https://link.aps.org/doi/10.1103/PhysRevLett.107.250401}.

\bibitem[Wang et~al.(2013)Wang, Ghobadi, Raeisi, and Simon]{Wang2013}
Tian Wang, Roohollah Ghobadi, Sadegh Raeisi, and Christoph Simon.
\newblock Precision requirements for observing macroscopic quantum effects.
\newblock \emph{Phys. Rev. A}, 88:\penalty0 062114, Dec 2013.
\newblock \doi{10.1103/PhysRevA.88.062114}.
\newblock URL \url{https://link.aps.org/doi/10.1103/PhysRevA.88.062114}.

\bibitem[Jeong et~al.(2014)Jeong, Lim, and Kim]{Jeong2014}
Hyunseok Jeong, Youngrong Lim, and M.~S. Kim.
\newblock Coarsening measurement references and the quantum-to-classical
  transition.
\newblock \emph{Phys. Rev. Lett.}, 112:\penalty0 010402, Jan 2014.
\newblock \doi{10.1103/PhysRevLett.112.010402}.
\newblock URL \url{https://link.aps.org/doi/10.1103/PhysRevLett.112.010402}.

\bibitem[Park et~al.(2014)Park, Ji, Lee, and Nha]{Park2014}
Jiyong Park, Se-Wan Ji, Jaehak Lee, and Hyunchul Nha.
\newblock Gaussian states under coarse-grained continuous variable
  measurements.
\newblock \emph{Phys. Rev. A}, 89:\penalty0 042102, Apr 2014.
\newblock \doi{10.1103/PhysRevA.89.042102}.
\newblock URL \url{https://link.aps.org/doi/10.1103/PhysRevA.89.042102}.

\bibitem[Duarte et~al.(2017)Duarte, Carvalho, Bernardes, and de~Melo]{cris2017}
Cristhiano Duarte, Gabriel~Dias Carvalho, Nadja~K. Bernardes, and Fernando
  de~Melo.
\newblock Emerging dynamics arising from coarse-grained quantum systems.
\newblock \emph{Phys. Rev. A}, 96:\penalty0 032113, Sep 2017.
\newblock \doi{10.1103/PhysRevA.96.032113}.
\newblock URL \url{https://link.aps.org/doi/10.1103/PhysRevA.96.032113}.

\bibitem[Silva~Correia and de~Melo(2019)]{pedrinho}
Pedro Silva~Correia and Fernando de~Melo.
\newblock Spin-entanglement wave in a coarse-grained optical lattice.
\newblock \emph{Phys. Rev. A}, 100:\penalty0 022334, Aug 2019.
\newblock \doi{10.1103/PhysRevA.100.022334}.
\newblock URL \url{https://link.aps.org/doi/10.1103/PhysRevA.100.022334}.

\bibitem[Kabernik(2018)]{oleg}
Oleg Kabernik.
\newblock Quantum coarse graining, symmetries, and reducibility of dynamics.
\newblock \emph{Phys. Rev. A}, 97:\penalty0 052130, May 2018.
\newblock \doi{10.1103/PhysRevA.97.052130}.
\newblock URL \url{https://link.aps.org/doi/10.1103/PhysRevA.97.052130}.

\bibitem[Duarte(2020)]{cris2019}
Cristhiano Duarte.
\newblock Compatibility between agents as a tool for coarse-grained
  descriptions of quantum systems.
\newblock \emph{Journal of Physics A: Mathematical and Theoretical},
  53\penalty0 (39):\penalty0 395301, aug 2020.
\newblock \doi{10.1088/1751-8121/aba574}.
\newblock URL \url{https://doi.org/10.1088/1751-8121/aba574}.

\bibitem[Veeren and de~Melo(2020)]{isadora2020}
Isadora Veeren and Fernando de~Melo.
\newblock Entropic uncertainty relations and the quantum-to-classical
  transition.
\newblock \emph{Phys. Rev. A}, 102:\penalty0 022205, Aug 2020.
\newblock \doi{10.1103/PhysRevA.102.022205}.
\newblock URL \url{https://link.aps.org/doi/10.1103/PhysRevA.102.022205}.

\bibitem[Carvalho and Correia(2020)]{gabriel2020}
Gabriel~Dias Carvalho and Pedro~Silva Correia.
\newblock Decay of quantumness in a measurement process: Action of a
  coarse-graining channel.
\newblock \emph{Phys. Rev. A}, 102:\penalty0 032217, Sep 2020.
\newblock \doi{10.1103/PhysRevA.102.032217}.
\newblock URL \url{https://link.aps.org/doi/10.1103/PhysRevA.102.032217}.

\bibitem[Duarte et~al.(2020)Duarte, Amaral, Cunha, and Leifer]{cris2020}
Cristhiano Duarte, Barbara Amaral, Marcelo~Terra Cunha, and Matthew Leifer.
\newblock Investigating coarse-grainings and emergent quantum dynamics with
  four mathematical perspectives.
\newblock \emph{arXiv preprint arXiv:2011.10349}, 2020.
\newblock \doi{10.48550/arXiv.2011.10349}.
\newblock URL \url{https://doi.org/10.48550/arXiv.2011.10349}.

\bibitem[Pineda et~al.(2021)Pineda, Davalos, Viviescas, and
  Rosado]{carlospineda2021}
Carlos Pineda, David Davalos, Carlos Viviescas, and Antonio Rosado.
\newblock Fuzzy measurements and coarse graining in quantum many-body systems.
\newblock \emph{Phys. Rev. A}, 104:\penalty0 042218, Oct 2021.
\newblock \doi{10.1103/PhysRevA.104.042218}.
\newblock URL \url{https://link.aps.org/doi/10.1103/PhysRevA.104.042218}.

\bibitem[Correia et~al.(2021)Correia, Obando, Vallejos, and
  de~Melo]{correia2021}
Pedro~Silva Correia, Paola~Concha Obando, Ra\'ul~O. Vallejos, and Fernando
  de~Melo.
\newblock Macro-to-micro quantum mapping and the emergence of nonlinearity.
\newblock \emph{Phys. Rev. A}, 103:\penalty0 052210, May 2021.
\newblock \doi{10.1103/PhysRevA.103.052210}.
\newblock URL \url{https://link.aps.org/doi/10.1103/PhysRevA.103.052210}.

\bibitem[Gross and Bloch(2017)]{gross2017quantum}
Christian Gross and Immanuel Bloch.
\newblock Quantum simulations with ultracold atoms in optical lattices.
\newblock \emph{Science}, 357\penalty0 (6355):\penalty0 995--1001, 2017.
\newblock \doi{10.1126/science.aal3837}.
\newblock URL \url{https://doi.org/10.1126/science.aal3837}.

\bibitem[Bloch(2008)]{fig1bloch}
Immanuel Bloch.
\newblock Quantum coherence and entanglement with ultracold atoms in optical
  lattices.
\newblock \emph{Nature}, 453\penalty0 (7198):\penalty0 1016--1022, 2008.
\newblock \doi{10.1038/nature07126}.
\newblock URL \url{https://doi.org/10.1038/nature07126}.

\bibitem[Fukuhara et~al.(2015)Fukuhara, Hild, Zeiher, Schau\ss{}, Bloch,
  Endres, and Gross]{fukuhara}
Takeshi Fukuhara, Sebastian Hild, Johannes Zeiher, Peter Schau\ss{}, Immanuel
  Bloch, Manuel Endres, and Christian Gross.
\newblock Spatially resolved detection of a spin-entanglement wave in a
  bose-hubbard chain.
\newblock \emph{Phys. Rev. Lett.}, 115:\penalty0 035302, Jul 2015.
\newblock \doi{10.1103/PhysRevLett.115.035302}.
\newblock URL \url{https://link.aps.org/doi/10.1103/PhysRevLett.115.035302}.

\bibitem[Sherson et~al.(2010)Sherson, Weitenberg, Endres, Cheneau, Bloch, and
  Kuhr]{sherson}
Jacob~F Sherson, Christof Weitenberg, Manuel Endres, Marc Cheneau, Immanuel
  Bloch, and Stefan Kuhr.
\newblock Single-atom-resolved fluorescence imaging of an atomic mott
  insulator.
\newblock \emph{Nature}, 467\penalty0 (7311):\penalty0 68--72, 2010.
\newblock \doi{10.1038/nature09378}.
\newblock URL \url{https://doi.org/10.1038/nature09378}.

\bibitem[Duranthon and Di~Molfetta(2021)]{dimolfetta21}
O.~Duranthon and Giuseppe Di~Molfetta.
\newblock Coarse-grained quantum cellular automata.
\newblock \emph{Phys. Rev. A}, 103:\penalty0 032224, Mar 2021.
\newblock \doi{10.1103/PhysRevA.103.032224}.
\newblock URL \url{https://link.aps.org/doi/10.1103/PhysRevA.103.032224}.

\bibitem[Laliena(1999)]{laliena99}
Victor Laliena.
\newblock Effect of angular momentum conservation in the phase transitions of
  collapsing systems.
\newblock \emph{Phys. Rev. E}, 59:\penalty0 4786--4794, May 1999.
\newblock \doi{10.1103/PhysRevE.59.4786}.
\newblock URL \url{https://link.aps.org/doi/10.1103/PhysRevE.59.4786}.

\bibitem[Lakshminarayan et~al.(2008)Lakshminarayan, Tomsovic, Bohigas, and
  Majumdar]{laksh08}
Arul Lakshminarayan, Steven Tomsovic, Oriol Bohigas, and Satya~N. Majumdar.
\newblock Extreme statistics of complex random and quantum chaotic states.
\newblock \emph{Phys. Rev. Lett.}, 100:\penalty0 044103, Jan 2008.
\newblock \doi{10.1103/PhysRevLett.100.044103}.
\newblock URL \url{https://link.aps.org/doi/10.1103/PhysRevLett.100.044103}.

\bibitem[Altland and Simons(2010)]{altland}
Alexander Altland and Ben~D. Simons.
\newblock \emph{Condensed Matter Field Theory}.
\newblock Cambridge University Press, 2010.

\bibitem[Inc.()]{Mathematica}
Wolfram~Research{,} Inc.
\newblock Wolfram programming lab, {V}ersion 13.0.0.
\newblock URL \url{https://www.wolfram.com/programming-lab}.

\bibitem[Zyczkowski and Sommers(2001)]{zyczkowski01}
Karol Zyczkowski and Hans-Jürgen Sommers.
\newblock Induced measures in the space of mixed quantum states.
\newblock \emph{Journal of Physics A: Mathematical and General}, 34\penalty0
  (35):\penalty0 7111--7125, aug 2001.
\newblock \doi{10.1088/0305-4470/34/35/335}.
\newblock URL \url{https://doi.org/10.1088/0305-4470/34/35/335}.

\bibitem[Saideh et~al.(2015)Saideh, Ribeiro, Ferrini, Coudreau, Milman, and
  Keller]{ibrahim}
Ibrahim Saideh, A.~D. Ribeiro, Giulia Ferrini, Thomas Coudreau, P\'erola
  Milman, and Arne Keller.
\newblock General dichotomization procedure to provide qudit entanglement
  criteria.
\newblock \emph{Phys. Rev. A}, 92:\penalty0 052334, Nov 2015.
\newblock \doi{10.1103/PhysRevA.92.052334}.
\newblock URL \url{https://link.aps.org/doi/10.1103/PhysRevA.92.052334}.

\bibitem[Binder et~al.(2018)Binder, Correa, Gogolin, Anders, and
  Adesso]{binder2018QT}
Felix Binder, Luis~A Correa, Christian Gogolin, Janet Anders, and Gerardo
  Adesso.
\newblock \emph{Thermodynamics in the quantum regime}, volume 195.
\newblock Springer, 2018.

\bibitem[Campisi et~al.(2011)Campisi, H\"anggi, and
  Talkner]{Colloquium2011quantumFluctuation}
Michele Campisi, Peter H\"anggi, and Peter Talkner.
\newblock Colloquium: Quantum fluctuation relations: Foundations and
  applications.
\newblock \emph{Rev. Mod. Phys.}, 83:\penalty0 771--791, Jul 2011.
\newblock \doi{10.1103/RevModPhys.83.771}.
\newblock URL \url{https://link.aps.org/doi/10.1103/RevModPhys.83.771}.

\bibitem[Allahverdyan(2014)]{allahverdyan14}
A.~E. Allahverdyan.
\newblock Nonequilibrium quantum fluctuations of work.
\newblock \emph{Phys. Rev. E}, 90:\penalty0 032137, Sep 2014.
\newblock \doi{10.1103/PhysRevE.90.032137}.
\newblock URL \url{https://link.aps.org/doi/10.1103/PhysRevE.90.032137}.

\bibitem[Talkner et~al.(2007)Talkner, Lutz, and H\"anggi]{Talker2007}
Peter Talkner, Eric Lutz, and Peter H\"anggi.
\newblock Fluctuation theorems: Work is not an observable.
\newblock \emph{Phys. Rev. E}, 75:\penalty0 050102, May 2007.
\newblock \doi{10.1103/PhysRevE.75.050102}.
\newblock URL \url{https://link.aps.org/doi/10.1103/PhysRevE.75.050102}.

\bibitem[Caves et~al.(2002)Caves, Fuchs, and Schack]{Caves2002}
Carlton~M. Caves, Christopher~A. Fuchs, and R\"udiger Schack.
\newblock Quantum probabilities as bayesian probabilities.
\newblock \emph{Phys. Rev. A}, 65:\penalty0 022305, Jan 2002.
\newblock \doi{10.1103/PhysRevA.65.022305}.
\newblock URL \url{https://link.aps.org/doi/10.1103/PhysRevA.65.022305}.

\bibitem[Blume-Kohout(2010)]{Blume2010}
Robin Blume-Kohout.
\newblock Optimal, reliable estimation of quantum states.
\newblock \emph{New Journal of Physics}, 12\penalty0 (4):\penalty0 043034, apr
  2010.
\newblock \doi{10.1088/1367-2630/12/4/043034}.
\newblock URL \url{https://doi.org/10.1088/1367-2630/12/4/043034}.

\bibitem[Zyczkowski and Kus(1994)]{zyczkowski94}
K~Zyczkowski and M~Kus.
\newblock Random unitary matrices.
\newblock \emph{Journal of Physics A: Mathematical and General}, 27\penalty0
  (12):\penalty0 4235--4245, jun 1994.
\newblock \doi{10.1088/0305-4470/27/12/028}.
\newblock URL \url{https://doi.org/10.1088/0305-4470/27/12/028}.

\end{thebibliography}

\end{document}